\pdfoutput=1
\documentclass[11pt,twoside,a4paper,cmspaper,final,collab]{cms-tdr}
\def\svnVersion{62819d1}\def\svnDate{2024/10/06}\def\cmsCernNoTag{CERN-EP-2024-155}\def\cmsCernDate{\today}\def\cmsMessage{Published in Physics Letters B as \href{http://dx.doi.org/10.1016/j.physletb.2024.139044}{\doi{10.1016/j.physletb.2024.139044}.}}
\begin{document}\cmsNoteHeader{BPH-22-009}

\newlength\cmsFigWidth
\ifthenelse{\boolean{cms@external}}{\setlength\cmsFigWidth{0.95\columnwidth}}{\setlength\cmsFigWidth{0.45\textwidth}}
\newlength\cmsFigWide
\ifthenelse{\boolean{cms@external}}{\setlength\cmsFigWide{\columnwidth}}{\setlength\cmsFigWide{0.7\textwidth}}
\newlength\cmsFigOne
\ifthenelse{\boolean{cms@external}}{\setlength\cmsFigOne{1.07\cmsFigWidth}}{\setlength\cmsFigOne{\cmsFigWidth}}
\newlength\cmsTabSkip\setlength\cmsTabSkip{10pt}
\ifthenelse{\boolean{cms@external}}{\providecommand{\cmsLeft}{top\xspace}}{\providecommand{\cmsLeft}{left\xspace}}
\ifthenelse{\boolean{cms@external}}{\providecommand{\cmsRight}{bottom\xspace}}{\providecommand{\cmsRight}{right\xspace}}
\ifthenelse{\boolean{cms@external}}{\providecommand{\cmsTable}[1]{\resizebox{\columnwidth}{!}{#1}}}{\providecommand{\cmsTable}[1]{#1}}

\newcommand{\QQbar}{\ensuremath{\PQQ\PAQQ}\xspace}
\newcommand{\jpsi}{\JPsi}
\newcommand{\psip}{\ensuremath{\PGy\mathrm{(2S)}}\xspace}
\newcommand{\lth}{\ensuremath{\lambda_\vartheta}\xspace}
\newcommand{\lph}{\ensuremath{\lambda_\varphi}\xspace}
\newcommand{\ltp}{\ensuremath{\lambda_{\vartheta\varphi}}\xspace}
\newcommand{\lthHX}{\ensuremath{\lambda_\vartheta^\mathrm{HX}}\xspace}
\newcommand{\costh}{\ensuremath{\cos\vartheta_\mathrm{HX}}\xspace}
\newcommand{\abscosth}{\ensuremath{\abs{\cos\vartheta_\mathrm{HX}}}\xspace}
\newcommand{\phiHX}{\ensuremath{\varphi_\mathrm{HX}}\xspace}
\newcommand{\tinynegspace}{{\kern-1pt}}
\newcommand{\oneSzero}{\ensuremath{{}^1{\tinynegspace}S_0^{[8]}}\xspace}
\newcommand{\threeSone}{\ensuremath{{}^3{\tinynegspace}S_1^{[8]}}\xspace}
\newcommand{\threePJ}{\ensuremath{{}^3{\tinynegspace}P_J^{[8]}}\xspace}
\newcommand{\threeSonePlusthreePJ}{%
\ensuremath{{}^3{\tinynegspace}S_1^{[8]} + \kappa ~ {}^3{\tinynegspace}P_J^{[8]}}\xspace}
\newcommand{\CNP}{\ensuremath{C_\mathrm{NP}}\xspace}
\newcommand{\CPR}{\ensuremath{C_\mathrm{PR}}\xspace}
\newcommand{\psiB}{\ensuremath{\psi_{\PB}}\xspace}
\newcommand{\fNPSpsiB}{\ensuremath{f^\mathrm{NPS}_{\psiB}}\xspace}
\newcommand{\fNPSCNP}{\ensuremath{f^\mathrm{NPS}_\mathrm{\CNP}}\xspace}
\newcommand{\fPRSpsiP}{\ensuremath{f^\mathrm{PRS}_{\psi_\mathrm{P}}}\xspace}
\newcommand{\fPRSpsiB}{\ensuremath{f^\mathrm{PRS}_{\psiB}}\xspace}
\newcommand{\fPRSCPR}{\ensuremath{f^\mathrm{PRS}_\mathrm{\CPR}}\xspace}

\cmsNoteHeader{BPH-22-009}

\title{Measurement of the polarizations of prompt and non-prompt \texorpdfstring{\jpsi and \psip}{J/psi and psi2S} mesons produced in $\Pp\Pp$ collisions at \texorpdfstring{$\sqrt{s} = 13$\TeV}{sqrt(s) = 13 TeV}}

\date{\today}

\abstract{The polarizations of prompt and non-prompt \jpsi and \psip mesons are measured in proton-proton collisions at $\sqrt{s} = 13$\TeV, using data samples collected by the CMS experiment in 2017 and 2018, corresponding to a total integrated luminosity of 103.3\fbinv. Based on the analysis of the dimuon decay angular distributions in the helicity frame, the polar anisotropy, \lth, is measured as a function of the transverse momentum, \pt, of the charmonium states, in the 25--120 and 20--100\GeV ranges for the \jpsi and \psip, respectively. The non-prompt polarizations agree with predictions based on the hypothesis that, for $\pt \gtrsim 25\GeV$, the non-prompt \jpsi and \psip are predominantly produced in two-body \PB meson decays. The prompt results clearly exclude strong transverse polarizations, even for \pt exceeding 30 times the \jpsi mass, where \lth tends to an asymptotic value around 0.3. Taken together with previous measurements, by CMS and LHCb at $\sqrt{s} = 7$\TeV, the prompt polarizations show a significant variation with \pt, at low \pt.}

\hypersetup{%
pdfauthor={CMS Collaboration},%
pdftitle={Measurement of the polarizations of prompt and non-prompt J/psi and psi2S mesons produced in pp collisions at sqrts = 13 TeV},%
pdfsubject={CMS},
pdfkeywords={CMS, quarkonium production, quarkonium polarization}}

\maketitle

\section{Introduction}

Charmonium and bottomonium production provides an ideal case study 
for the understanding of hadron formation 
in quantum chromodynamics (QCD)~\cite{QWG:2004,Brambilla}. 
Its theoretical description is based on the generally agreed assumption 
that the charm and beauty quarks 
(the heaviest ones capable of forming bound states)
are heavy enough to allow the factorization of short- and long-distance effects.
Within the non-relativistic QCD (NRQCD) framework~\cite{NRQCD}, in particular, 
perturbative QCD computations provide the production cross sections of the \QQbar heavy-quark pair
(the ``short-distance coefficients", SDCs),
while the non-perturbative evolution of the \QQbar state to the observed meson 
(hadronization) is described by phenomenological parameters
(the ``long-distance matrix elements", LDMEs), 
determined from fits to experimental data.
Other theoretical approaches have been considered, 
such as the colour-singlet model (CSM)~\cite{Baier:1981uk,Lansberg-HP08} 
and the colour-evaporation model (CEM)~\cite{CEM_BargerPLB,CEM_BargerZPC}.
These theoretical models differ in the choice and classification of the allowed pre-resonance states. 
The NRQCD approach foresees the contribution of all possible spin, $S$, 
orbital angular momentum, $L$, total angular momentum, $J$, 
and colour ($c = 1$ or 8) configurations, $\QQbar(^{2S+1}L_J^{[c]})$, 
organized in an expansion in powers of the relative \QQbar velocity
so that only a small number of leading and sub-leading terms remain quantitatively important.
Instead, the CSM considers that the final-state hadron can only result from a
colour-neutral (singlet) pre-resonance state having the same quantum numbers,
while the CEM is built upon the assumption that 
the probability of forming a specific quarkonium state is independent 
of its kinematics and spin, as well as of the production process.

According to the perturbative calculations of the relevant partonic processes, 
the observable transverse momentum (\pt) distributions of the produced quarkonium meson
depend significantly on the angular momentum quantum state 
of the unobservable \QQbar pre-resonance. 
Moreover, the contributing short-distance processes 
(one, few, or many, depending on the model)
are scaled by long-distance weights that further depend on 
the angular momentum quantum numbers of each pre-resonance. 
Therefore, by foreseeing different pre-resonance mixtures, 
the CSM, CEM, and NRQCD predict, in general, 
different distributions for \pt and other kinematic variables.
The polarization of the quarkonium state is, however, 
the observable that most directly reflects the mixture of $S, L, J$ configurations 
(and polarizations) of the contributing pre-resonance states,
as it can be understood independently of any perturbative calculation.
Consequently, polarization measurements provide particularly straightforward information
regarding the details of the hadronization models.
The polarizations of five vector ($J^{PC} = 1^{--}$) quarkonia,
\jpsi, \psip, $\Upsilon$(1S), $\Upsilon$(2S) and $\Upsilon$(3S),
have been measured, as functions of \pt,
at the Fermilab Tevatron~\cite{CDF:psiPol,CDF:UpsilonPol}
and the CERN LHC~\cite{BPH-13-003,LHCb_psiPol,LHCb_psi2SPol,BPH-11-023,LHCb_UpsilonPol}.
These measurements have been considered in several phenomenological studies,
including analyses based on the
NRQCD~\cite{Butenschoen:2012px,ChaoPRL,Gong:2012ug,Butenschoen:2012qr,
Faccioli:PLB736,Bodwin:2015iua,Faccioli2018}
and CEM~\cite{iCEM-pol} approaches.
Measurements of inclusive \jpsi polarizations, 
where the contribution from decays of \Pb hadrons is not subtracted,
have been reported by the PHENIX, STAR, and ALICE Collaborations~\cite{PHENIX:2009ghc, STAR:2020igu, ALICE:2011gej}.

Within the precision of the previous CMS measurements~\cite{BPH-13-003,BPH-13-001},
no significant deviation is seen with respect to the ``unpolarized scenario",
where the directly (\ie, excluding feed-down decays) produced \jpsi mesons have zero and \pt-independent polarization~\cite{CPT}.
The straightforward interpretation of that scenario would be that 
\jpsi production is dominated by the (unpolarized) \oneSzero colour-octet state,
an option not naturally foreseen by NRQCD, 
where the LDMEs of the strongly polarized \threeSone and \threePJ octet contributions 
and of the \oneSzero term are expected to have similar magnitudes, 
leading to a significant \pt dependence of the polarization. 
In principle, one might think that sufficiently precise \pt-differential cross section measurements
would be able to discriminate between those two scenarios,
one where the \oneSzero term dominates and 
the other also including significant \threeSone and \threePJ contributions.
In practice, however, this is not the case, because of two facts~\cite{Faccioli2019}. 
First, an accidental degeneracy makes the shape of a certain combination 
of the \threeSone and \threePJ \pt distributions indistinguishable from the \oneSzero one.
Second, both the shape of that combination and the shape of the \oneSzero term 
are very similar to the measured \jpsi \pt distribution, 
indicating that reality is critically close to the degeneracy condition. 
Therefore, the observed \pt distribution has a weak resolving power 
on the participating processes. 
In fact, fits trying to determine the relative process contributions, \ie the LDMEs, 
using only the measured differential cross sections as constraints 
can lead to ambiguous results~\cite{Faccioli:PLB736}.
The polarization measurement provides a completely independent and sensitive source of information:
any $\threeSone + \threePJ$ combination leads to a recognizable \pt-dependent polarization, 
changing monotonically from longitudinal to transverse as \pt increases.
Therefore, a sufficiently precise polarization measurement, performed over a wide-enough \pt range, 
should be able to reveal the relative contributions of the differently polarized colour octet terms.

This Letter reports a new measurement of the prompt \jpsi and \psip polarizations,
in proton-proton ($\Pp\Pp$) collisions at a centre-of-mass energy of 13\TeV,
based on data collected by the CMS experiment in 2017 and 2018, 
corresponding to a total integrated luminosity of 103.3\fbinv. 
This event sample is much larger than the one collected in 2011, at 7\TeV, 
which was used for the previous measurement~\cite{BPH-13-003}.
The results are sufficiently accurate to provide
the sensitivity needed to evaluate the relative proportions of 
the three colour octet channels.
We also present the first LHC measurement of the polarizations of non-prompt \jpsi and \psip mesons,
predominantly produced in decays of \PB mesons,
with a small contribution from decays of other \Pb hadrons.
This result provides an independent probe of the charmonium formation mechanism
and of its composition in terms of singlet and octet contributions.
Besides the directly produced component, the prompt \jpsi yield 
includes fractions from ``feed-down decays" of heavier charmonium states:
around 8\% from \psip decays and 
25\% from $\chi_{\PQc}$ decays~\cite{Faccioli-feeddown}.

The average polarizations of $J^{PC}=1^{--}$ quarkonia are usually determined 
by measuring the angular distributions of the positively charged muons emitted in the decay of the mesons,
which have the general observable form~\cite{Faccioli-PRL-FrameInv,Faccioli-PRD-FrameInv}
\ifthenelse{\boolean{cms@external}}
{
\begin{equation}
\begin{split}
& W( \cos \vartheta, \varphi \; | \; \vec{\lambda} ) \,
 = \, \frac{3}{4 \pi (3 + \lth)} \, \left(1 + \lth \cos^2 \vartheta \right.\\
& \left. +\, \lph \sin^2 \vartheta \cos 2 \varphi + \ltp \sin 2 \vartheta \cos \varphi \right),
\end{split}
\label{eq:ang_distr_general}
\end{equation}
}
{
\begin{equation}
W( \cos \vartheta, \varphi \; | \; \vec{\lambda} ) \,
 = \, \frac{3}{4 \pi (3 + \lth)} \, (1 + \lth \cos^2 \vartheta
 + \lph \sin^2 \vartheta \cos 2 \varphi + \ltp \sin 2 \vartheta \cos \varphi ),
\label{eq:ang_distr_general}
\end{equation}
}
where $\vartheta$ and $\varphi$ are the polar and azimuthal angles of the muon 
in the quarkonium rest frame with respect to, respectively, 
a suitably defined polarization axis $z$ and 
the plane containing the momenta of the colliding beams and of the quarkonium~\cite{PPHEP}.
The shape of the decay angular distribution is defined by 
the polarization parameters \lth, \lph, and \ltp.
Depending on the chosen polarization frame, 
not all three parameters contain equally significant and/or independent physical information, 
as is extensively discussed in Ref.~\cite{PPHEP}. 
In particular, in a domain where the laboratory momentum is always much larger 
than the particle mass $M$, it can be safely assumed 
(as seen, for example, in Drell--Yan and Z boson production measurements 
in the limit of very high $\pt/M$~\cite{CMS_Z_A0, ATLAS_Z_A0})
that azimuthal anisotropies become negligible in the centre-of-mass helicity frame (HX),
where the $z$ axis coincides with the particle momentum direction in
the centre-of-mass frame of the colliding hadrons~\cite{PPHEP}.
Moreover, at mid-rapidity and in the limit of high \pt, 
relevant for the present measurement, the shape 
of the muon pair acceptance as a function of $\cos\vartheta_\mathrm{HX}$ 
becomes essentially independent of $\varphi_\mathrm{HX}$ 
so that Eq.~(\ref{eq:ang_distr_general}) can be integrated over the azimuthal decay angle 
and the analysis can focus on the simpler one-parameter distribution 
\begin{equation}
W(\cos\vartheta_\mathrm{HX}) \propto 1+\lthHX \cos^2 \vartheta_\mathrm{HX}, 
\label{eq:W}
\end{equation}
without the risk of being biased by neglected correlation effects.
In this phase space window (high \pt and mid-rapidity)
the perpendicular helicity frame~\cite{PX-frame} is indistinguishable from the HX frame.
On the other hand, the Collins--Soper (CS) frame~\cite{CS-frame} is unsuitable because 
it is not possible to integrate the 
$\cos\vartheta_\mathrm{CS}$ vs.\ $\varphi_\mathrm{CS}$ acceptance map 
over the azimuthal decay angle without including zero-acceptance domains, 
so that a two-dimensional analysis becomes necessary~\cite{PPHEP}.

\section{Apparatus, event samples, and selection criteria}

The CMS apparatus is a multipurpose detector~\cite{CMS:2008xjf}
designed to trigger on and identify electrons, muons, photons,
and (charged and neutral) hadrons~\cite{CMS:2020uim,CMS:2018rym,CMS:2014pgm}.
A superconducting solenoid of 6\unit{m} internal diameter provides a magnetic field of 3.8\unit{T}.
Within the solenoid volume are the silicon pixel and strip tracker,
a crystal electromagnetic calorimeter, and a brass and scintillator hadron calorimeter.
Muons are measured in gas-ionization detectors embedded in the steel flux-return yoke outside the solenoid.
Events of interest are selected using a two-tiered trigger system.
The first level, composed of custom hardware processors,
uses information from the calorimeters and muon detectors to select events
at a rate of 100\unit{kHz} within a fixed latency of 4\mus~\cite{CMS:2020cmk}.
The second level,
consisting of a farm of processors running a faster version of the full event reconstruction software,
reduces the rate to around 1\unit{kHz}, before data storage~\cite{CMS:2016ngn}.

The event samples used in the analysis were collected in 2017 and 2018, 
with integrated luminosities of 42.0 and 61.3\fbinv, respectively~\cite{CMS-PAS-LUM-17,CMS-PAS-LUM-18}.
The events were selected by two dimuon triggers, 
requiring an opposite-sign muon pair with the invariant mass
in the ranges $2.9 < M < 3.33$\GeV for the \jpsi case and 3.35--4.05\GeV for the \psip case.
The distance of closest approach between the two muons must be smaller than 0.5\cm and
a fit of the positions and momenta of the two muons to a common vertex
(``dimuon vertex fit") must have a $\chi^2$ probability larger than 0.5\%.
In addition, to reduce the trigger rate, the dimuon \pt must be larger than 24.9 and 17.9\GeV 
for the \jpsi and \psip events, respectively.
No explicit \pt\ requirement was imposed on the individual muons at the trigger level.
The dimuon rapidity is restricted to $\abs{y} < 1.25$,
where the muon momentum is measured with the best resolution.

The reconstructed data were processed ensuring that both reconstructed muons 
must match, in pseudorapidity ($\eta$) and azimuthal angle, 
those that triggered the detector readout.
Both muon tracks must have more than five hits in the silicon tracker, 
at least one of them being in a pixel detector layer.
They must also fulfill other (``soft-muon") identification requirements~\cite{CMS:2018rym},
which include the (loose) matching between the track reconstructed in the silicon tracker
and the one reconstructed in the muon detectors.
The two muons of the selected events must each have 
$\pt > 5.6$\GeV, $\abs{\eta} < 1.4$, 
and a dimuon vertex fit $\chi^2$ probability larger than 1\%.
The single muon \pt requirement ensures that all selected muons are in the plateau region of the
detection efficiency, so that the efficiency variations are smaller than 5\%.
The polarizations are measured 
in the \pt ranges 25--120\GeV (in 19 bins) and 20--100\GeV (in 8 bins)
for the \jpsi and \psip mesons, respectively.
The results are obtained in the dimuon rapidity window $\abs{y} < 1.2$.
The dimuon mass distributions,
studied in the 2.92--3.28 and \mbox{3.4--4.0}\GeV ranges
for the \jpsi and \psip cases, respectively,
provide the information needed to separate the signal contributions
(dimuons from the \jpsi and \psip decays) 
from the underlying continuum background, 
composed of muon pairs resulting from other processes, 
such as decays of heavy-flavour hadrons.

\begin{figure}[ht!]
\centering
\includegraphics[width=\cmsFigOne]{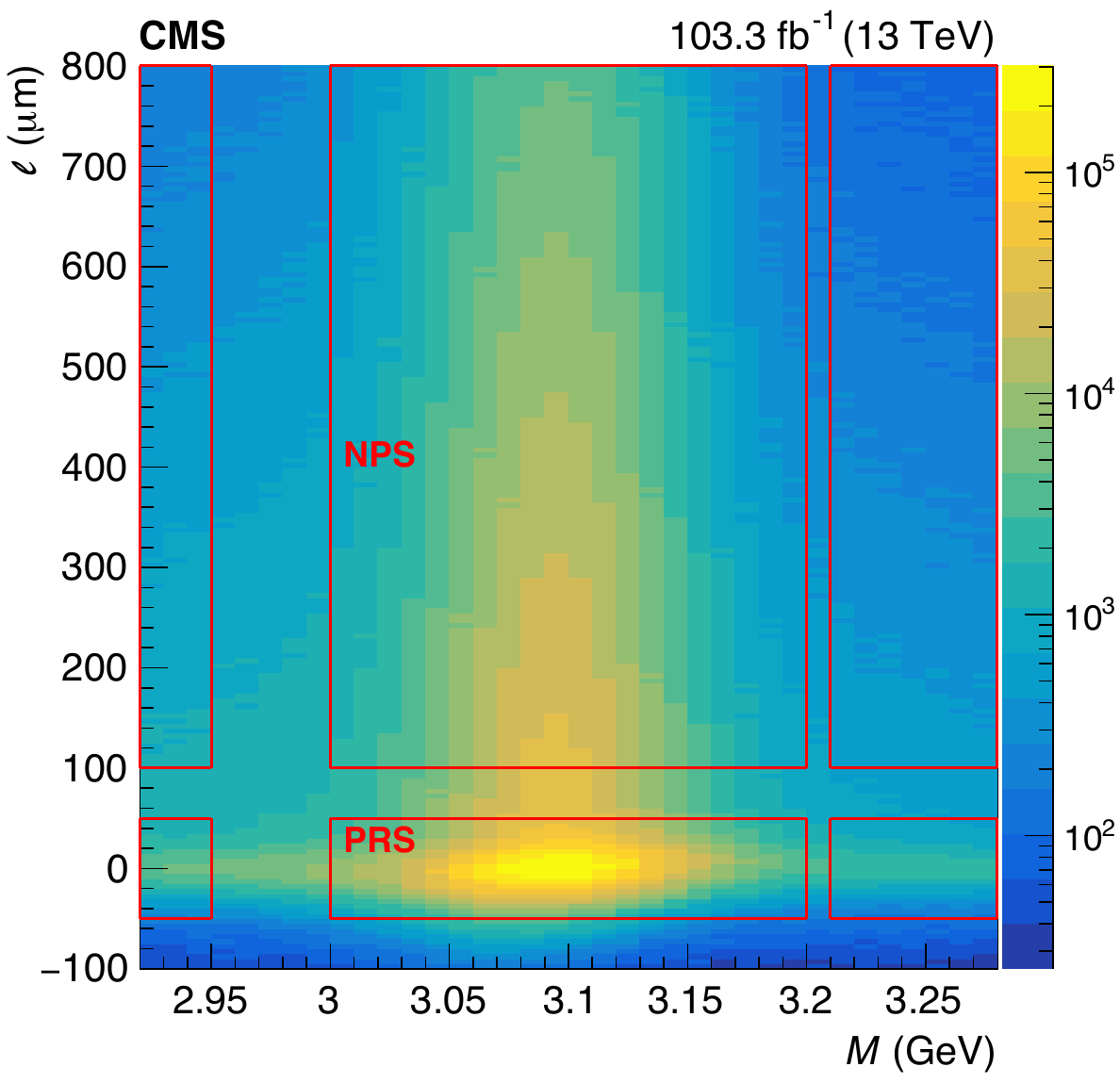}
\includegraphics[width=\cmsFigOne]{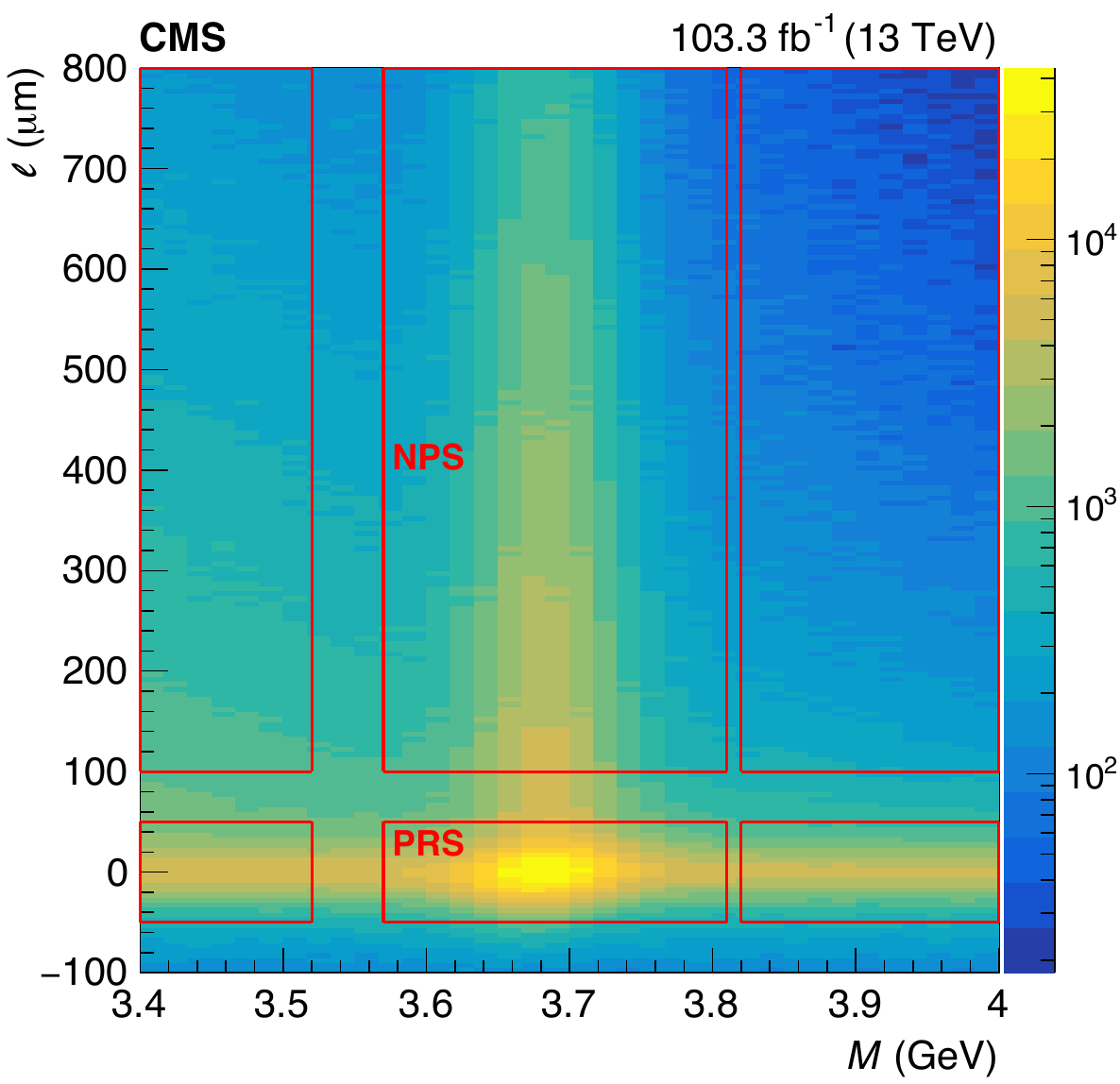}
\caption{Measured dimuon decay length vs.\ mass distributions
for the \jpsi (\cmsLeft) and \psip (\cmsRight) samples,
showing the rectangular regions used in the analysis.
The prompt and non-prompt signal regions are labeled PRS and NPS, respectively, 
while the remaining regions are non-signal (sideband) regions.}
\label{fig:ctau_vs_mass_map}
\end{figure}

For each of the two charmonium states and each of the \pt bins, 
the \abscosth distributions are measured in six independent event samples,
defined by three ranges in the dimuon mass (signal window and two sidebands)
and two in the dimuon pseudo-proper decay length (prompt and non-prompt).
The six windows are presented in Fig.~\ref{fig:ctau_vs_mass_map}, 
for the \jpsi and \psip analyses.
The dimuon pseudo-proper decay length~\cite{BPH-10-002}, 
abbreviated as ``decay length" in the remainder of this Letter,
is defined as $\ell = M \vec{L}_{xy} \cdot \vec{\pt} / \pt^2$, 
where $\vec{L}_{xy}$ is the displacement in the transverse plane 
between the primary vertex and the dimuon production vertex;
it is measured with a resolution around 25\mum.
The primary vertex is selected among all reconstructed $\Pp\Pp$ collision vertices in the event 
as the one closest to the line extrapolating the dimuon momentum back to the beam line 
and its position is determined by fitting all tracks associated with the vertex 
other than the two selected muons.
In the following, 
we use the labels PRS (``prompt signal") and NPS (``non-prompt signal")
to identify the dimuon mass vs.\ decay length two-dimensional windows
used to measure the prompt ($\abs{\ell} < 50\,\mu$m) 
and non-prompt ($100 < \ell < 800\,\mu$m)
\jpsi (3.0--3.2\GeV) and \psip (3.57--3.81\GeV) meson polarizations.
We also use PR and NP as subscripts to denote the event samples 
with $\abs{\ell} < 50\,\mu$m and $100 < \ell < 800\,\mu$m, respectively.
The \jpsi analysis uses 14.7~M and 10.9~M events in the PRS and NPS windows, respectively;
the corresponding numbers for the \psip analysis are 2.1~M and 1.4~M.
The other four windows are mass sidebands 
used for the subtraction of the continuum dimuon background:
2.92--2.95 and 3.21--3.28\GeV for the \jpsi analysis, and 
3.4--3.52 and 3.82--4.0\GeV for the \psip analysis.

The detection acceptance and efficiency effects are evaluated through detailed 
Monte Carlo (MC) simulations of the entire data collection and reconstruction chain, 
from the trigger step to the offline event selection.
The events are generated assuming unpolarized production, 
so that any non-flat trends seen in the reconstructed distributions 
are caused by the convolution of the detection effects.
The \jpsi and \psip event samples are generated with the 
\PYTHIA\,8.240 event generator~\cite{Pythia}.
The emitted muons undergo final-state radiation, 
generated through the \PHOTOS\,3.61 package~\cite{PHOTOS2}.
The simulation includes effects from multiple $\Pp\Pp$ interactions 
in the same or nearby bunch crossings, 
with a multiplicity distribution tuned to match the data.
The simulated events are then processed through a detailed simulation of the CMS detector,
based on the \GEANTfour package~\cite{Geant4}, 
using the same trigger and reconstruction algorithms as used 
to collect and process the data; 
they also need to pass the same selection criteria. 
The simulated samples are independently generated for each of the two data-taking years and 
several validation checks were performed to ensure that they 
reliably reproduce the running conditions of the experiment during those periods.
The polarization measurement is insensitive to the acceptance and efficiency magnitudes;
only their variation with \abscosth, in each \pt bin, is relevant.
The detection acceptance and efficiency is approximately flat with \costh,
except close to the edge of the covered \abscosth window, where it drops.
The coverage in \abscosth
is determined by the $\pt > 5.6$\GeV requirement on the individual muons and
increases from $\abscosth < 0.5$ to $\abscosth < 0.9$,
between the lowest and highest dimuon \pt bins of the analysis.

\section{Dimuon mass and decay length analysis}

\subsection{Analysis overview}

The polarizations measured in this analysis are reported using the polar anisotropy parameter \lthHX,
extracted by fitting the \abscosth distributions with Eq.~(\ref{eq:W});
in the following, for simplicity, we drop the superscript HX.
We start by describing the measurement of the polarizations of the non-prompt \jpsi and \psip mesons.
While a small fraction of those mesons is produced in decays of \Pb baryons, 
they are predominantly the daughters of \PB mesons and, hence, 
we denote them as \psiB.
The dimuons contributing to the \abscosth distribution in the non-prompt signal mass region for a given \pt bin, 
$\text{NPS}(\abscosth,\pt)$,
come either from decays of \psiB mesons
or from non-prompt continuum background processes, \CNP,
\ifthenelse{\boolean{cms@external}}
{
\begin{equation}
\begin{split}
&\text{NPS}(\abscosth,\pt)
= \fNPSpsiB(\pt) \, \psiB(\abscosth,\pt) \\
&+ \fNPSCNP(\pt) \, \CNP(\abscosth,\pt),
\end{split}
\label{eq:psi_NP}
\end{equation}
}
{
\begin{equation}
\begin{split}
\text{NPS}(\abscosth,\pt)
&= \fNPSpsiB(\pt) \, \psiB(\abscosth,\pt) \\
&+ \fNPSCNP(\pt) \, \CNP(\abscosth,\pt),
\end{split}
\label{eq:psi_NP}
\end{equation}
}
with $\fNPSpsiB(\pt) = 1 - \fNPSCNP(\pt)$.

Similarly, the polarizations of the prompt \jpsi and \psip mesons, $\psi_\mathrm{P}$,
can also be measured, in each \pt bin, using the \abscosth distributions
of the dimuons in the prompt signal region, $\text{PRS}(\abscosth,\pt)$.
However, besides the dimuons from prompt charmonia
and from continuum background processes, \CPR,
the PRS window also includes events from 
charmonia produced in decays of short-lived \PB mesons, of $\abs{\ell} < 50\,\mu$m,
\ifthenelse{\boolean{cms@external}}
{
\begin{equation}
\begin{split}
&\text{PRS}(\abscosth,\pt) 
= \fPRSpsiP(\pt) \, \psi_\mathrm{P}(\abscosth,\pt) \\
& + \fPRSCPR(\pt) \, \CPR(\abscosth,\pt) \\
& + \fPRSpsiB(\pt) \, \psiB(\abscosth,\pt),
\end{split}
\label{eq:psi_PR}
\end{equation}
}
{
\begin{equation}
\begin{split}
\text{PRS}(\abscosth,\pt) 
&= \fPRSpsiP(\pt) \, \psi_\mathrm{P}(\abscosth,\pt) \\
&+ \fPRSCPR(\pt) \, \CPR(\abscosth,\pt)
+ \fPRSpsiB(\pt) \, \psiB(\abscosth,\pt),
\end{split}
\label{eq:psi_PR}
\end{equation}
}
with $\fPRSpsiP(\pt) = 1 - \fPRSCPR(\pt) - \fPRSpsiB(\pt)$.

The fractions of continuum muon pairs in the signal windows,
\fNPSCNP and \fPRSCPR,
are determined from fits to the dimuon mass distributions,
while the fractions of charmonia from \PB meson decays in the prompt signal window, 
\fPRSpsiB, 
are obtained by fitting the dimuon decay length distributions, 
accounting for the existence of non-prompt continuum muon pairs.
The continuum dimuons in the PR region include prompt Drell--Yan dimuons
and several combinations of muons produced in decays 
of pions, kaons, \PD mesons, and \PB mesons.
The remainder of this section explains how the three fractions are measured,
for each of the two charmonium states and as functions of \pt, and reports the results,
which are then used in the measurement of the polarizations.

\subsection{Background fractions in the non-prompt charmonium signal regions}

\begin{figure}[ht!]
\centering
\includegraphics[width=\cmsFigWidth]{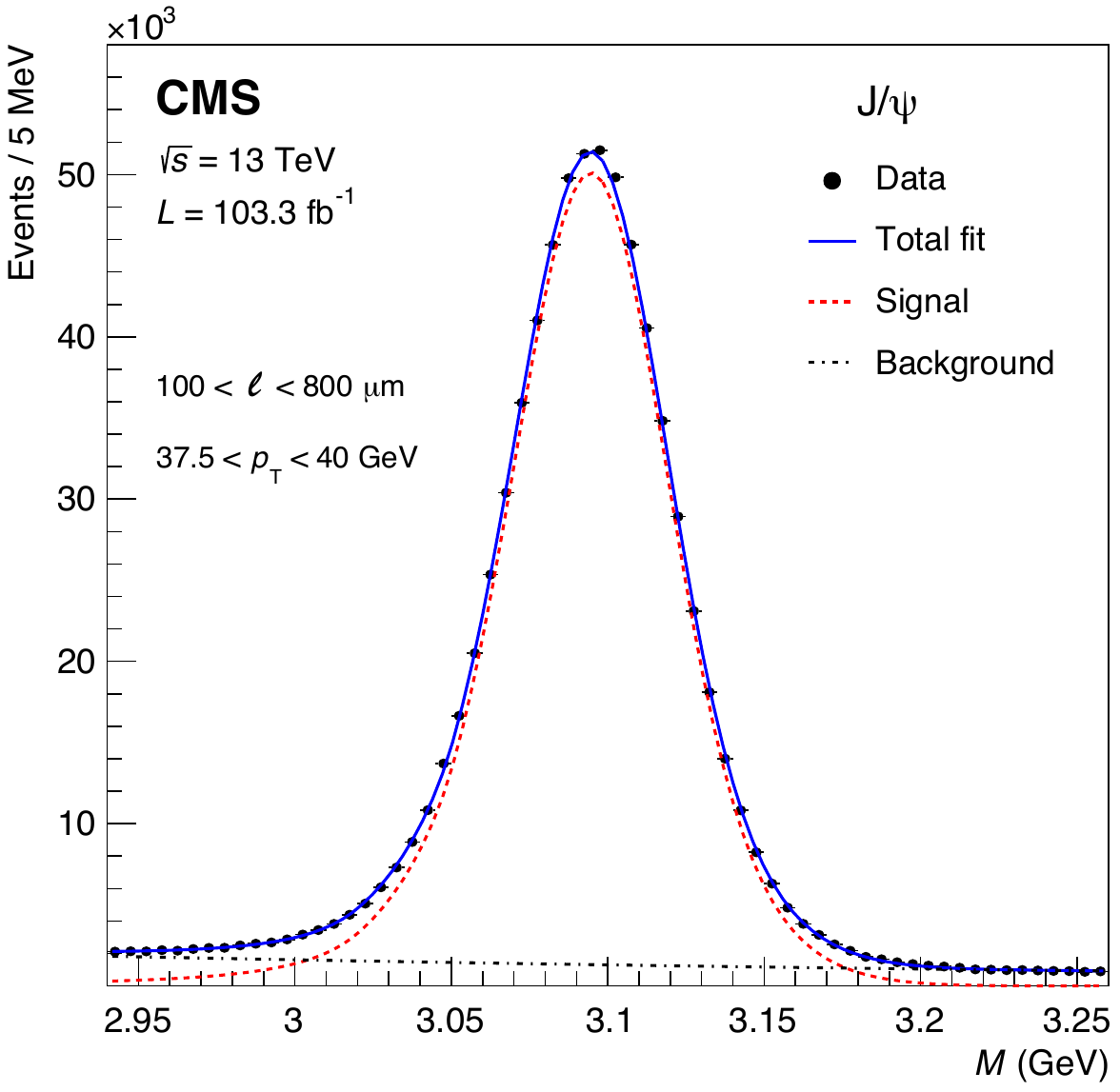}
\includegraphics[width=\cmsFigWidth]{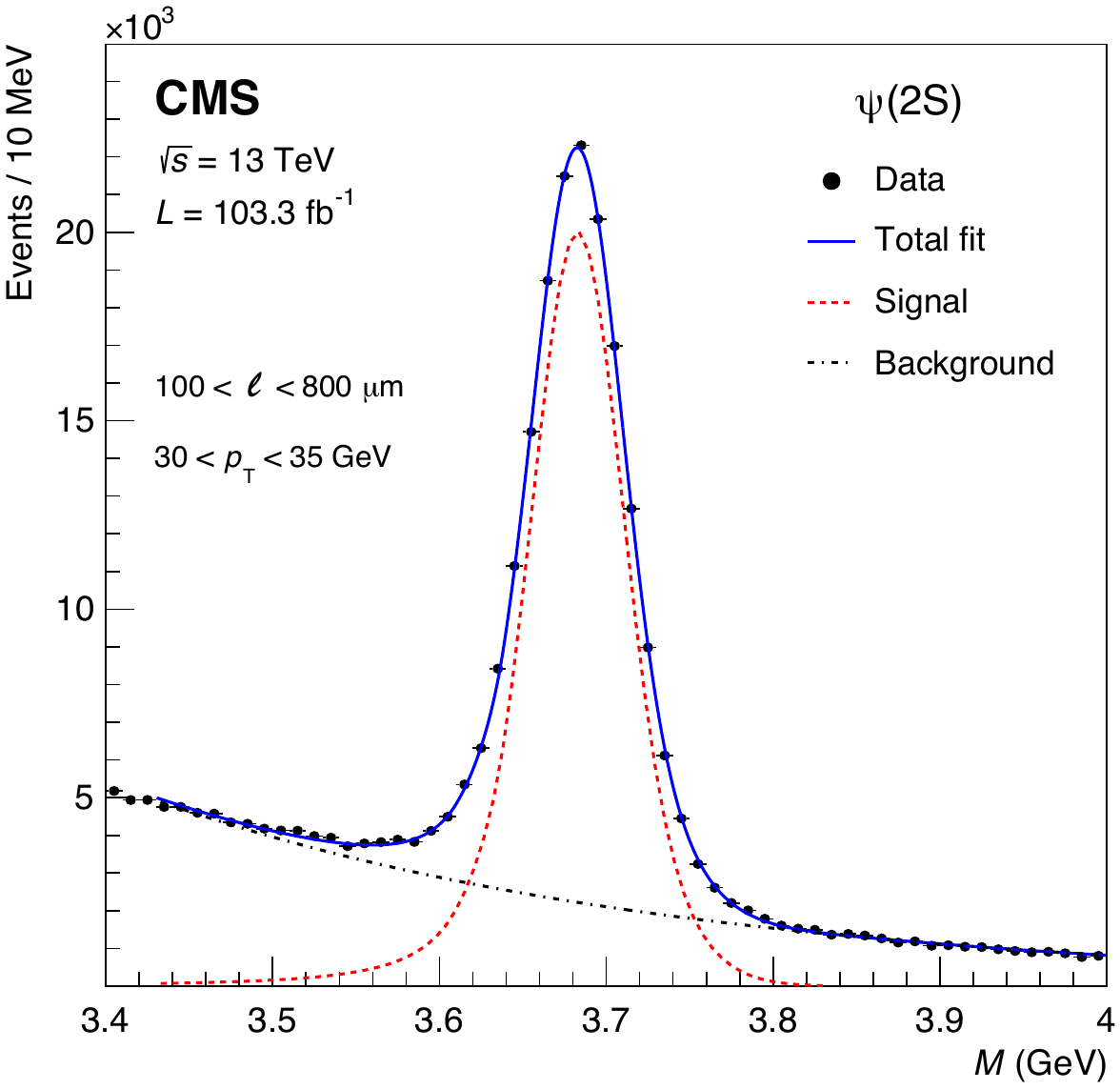}
\caption{Dimuon mass distributions measured for
the non-prompt \jpsi (\cmsLeft) and \psip (\cmsRight) event samples,
in the mentioned \pt bins.
The total fit function (blue), 
the sum of the two CB functions and, only in the \jpsi case, the Gaussian function (red),
and the background continuum (black) are also shown.}
\label{fig:NP-mass-fit}
\end{figure}

The fits of the non-prompt \jpsi and \psip dimuon mass distributions,
illustrated in the \cmsLeft and \cmsRight panels of Fig.~\ref{fig:NP-mass-fit}, respectively, 
are conceptually identical except for the functions describing the signal shape.
We start with the simpler \psip case.

The eight dimuon mass distributions
(corresponding to the eight \psip \pt bins)
are simultaneously fitted 
by the sum of two Crystal Ball (CB) functions~\cite{CB}, 
to describe the \psip line shape,
plus a decreasing exponential function,
to describe the underlying continuum background.
The two CB functions have common means ($\mu_1 = \mu_2$)
and tail parameters $n$ and $\alpha$, 
but independent widths ($\sigma_1$ and $\sigma_2$).
A good description of all of the \psip dimuon mass distributions is obtained 
with a relatively small number of free shape parameters 
by constraining their dependence on \pt 
following studies of the MC event samples validated on the measured data.
More specifically, the CB means and their relative proportions are independent of \pt,
as is the value of $\alpha$,
while the two widths increase linearly with \pt, with a common slope.
The tail parameter $n$ is fixed to 2.5, a value based on studies of simulated events;
given the strong correlation between $n$ and $\alpha$, 
it is reasonable to leave only one of them as a free parameter in the fit.
Figure~\ref{fig:NP-mass-fit}~(\cmsRight) shows
the measured non-prompt \psip dimuon mass distribution in a typical \pt bin.
The lines represent the result of the fit, 
which is simultaneously made to the eight mass distributions.

Since we have a much larger number of \jpsi events than \psip events,
we need to use a slightly more complex \jpsi signal fit model.
First, we noticed that the description of the peak line shape improves if we 
no longer impose that the two CB functions have a common mean; 
instead, we constrain $\mu_1$ to be independent of \pt and leave $\mu_2$ free in each \pt bin.
Second, we add a Gaussian function, 
with $\mu_1$ as mean and a normalization fixed 
from studies of the MC event samples, 
leading to a contribution of 3.5\% of the total \jpsi yield.
The Gaussian width, $\sigma_G$, 
is constrained to increase linearly with \pt with the same slope as $\sigma_1$ and $\sigma_2$.
Simultaneously fitting all 19 \jpsi dimuon mass distributions provides
a good description of the data.
Figure~\ref{fig:NP-mass-fit}~(\cmsLeft) shows
the non-prompt \jpsi dimuon mass distribution measured for one of the 19 \pt bins.

\begin{figure}[ht!]
\centering
\includegraphics[width=0.95\cmsFigWidth]{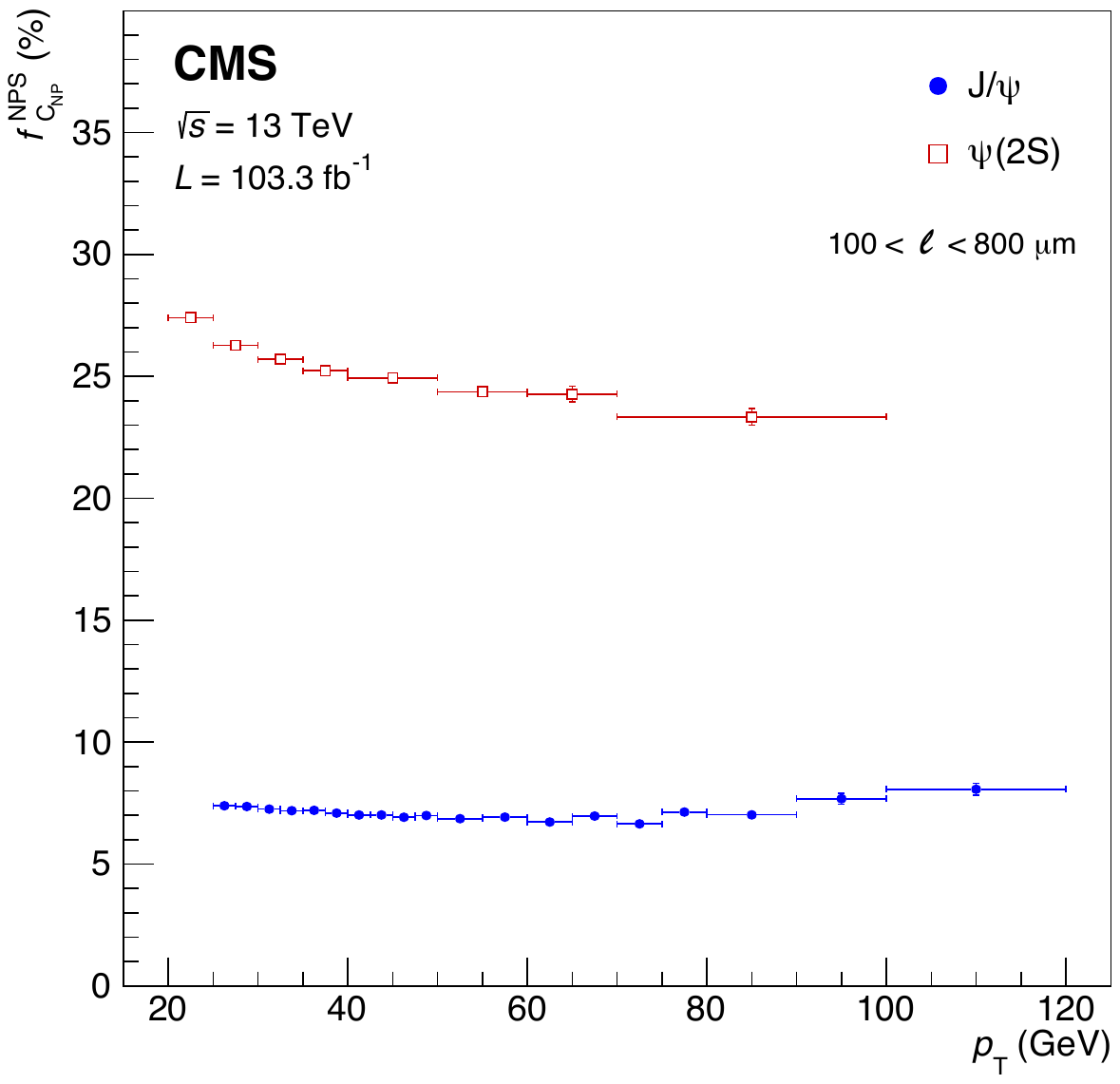}
\caption{Fraction of events in the NPS region due to continuum muon pairs,
versus \pt, for the non-prompt \jpsi and \psip events.}
\label{fig:NP_psi_Bg-fraction}
\end{figure}

The most important shape parameters obtained from the dimuon mass fits
(left free in each \pt bin)
are the amplitude and slope of the continuum exponential function
because they determine the fraction \fNPSCNP,
which is computed by integrating the fitted background function (in each \pt bin)
in the signal mass window (NPS) 
and dividing the result by the total number of events counted in that region.
Figure~\ref{fig:NP_psi_Bg-fraction} shows the obtained \fNPSCNP fractions, 
as functions of \pt.

\subsection{Background fractions in the prompt charmonium signal regions}

\begin{figure}[ht!]
\centering
\includegraphics[width=\cmsFigWidth]{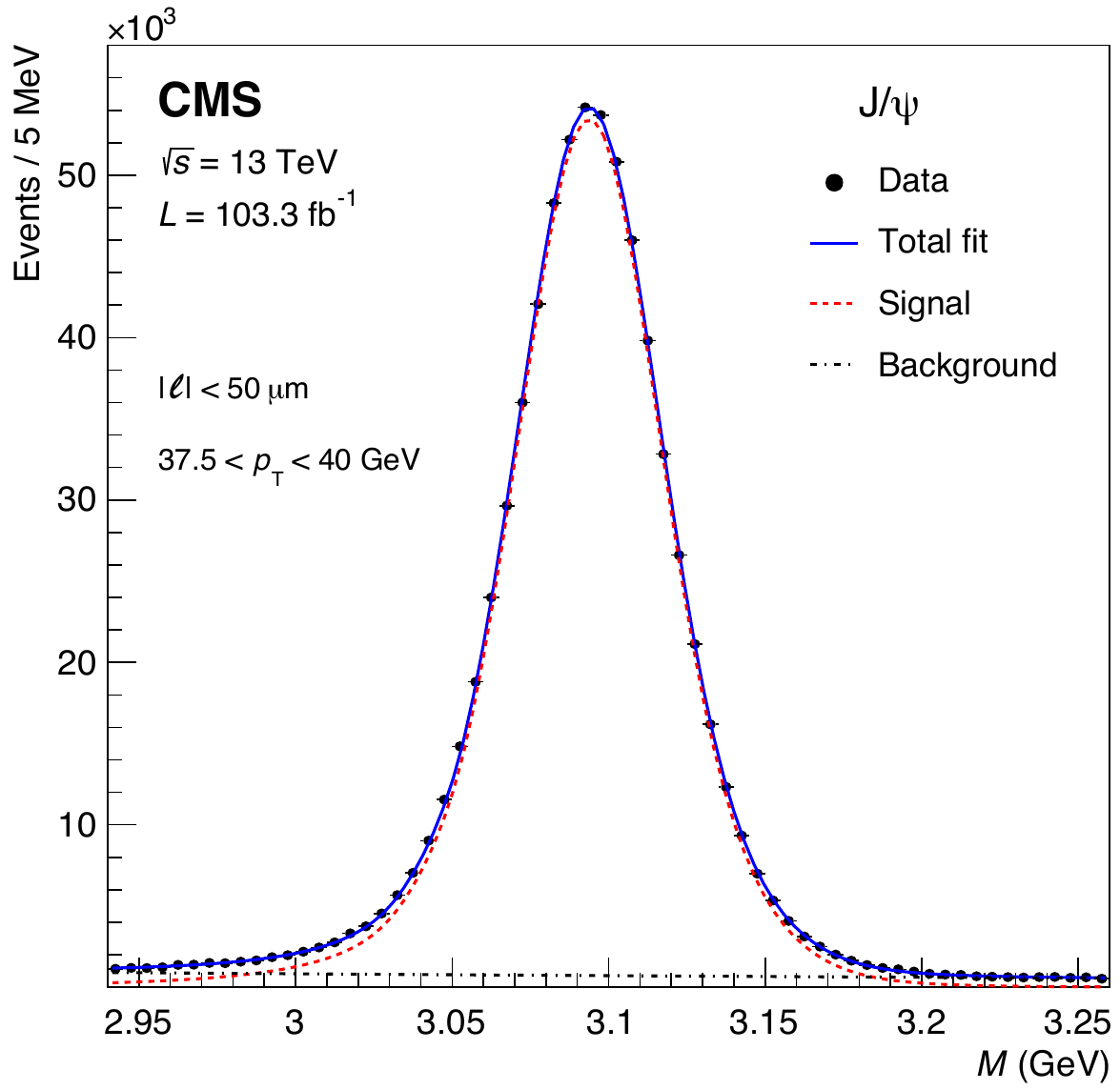}
\includegraphics[width=\cmsFigWidth]{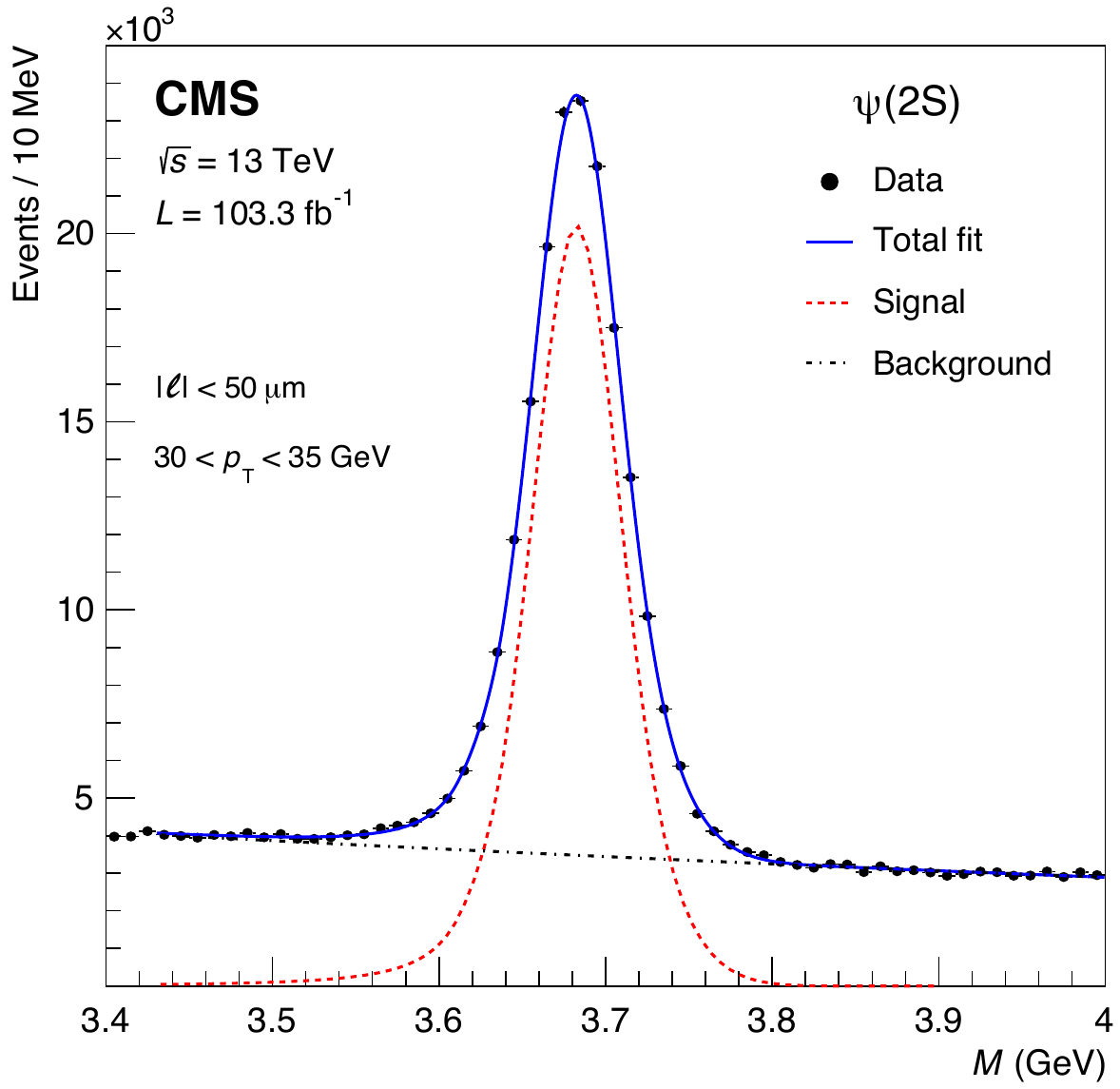}
\caption{Dimuon mass distributions measured for
the prompt \jpsi (\cmsLeft) and \psip (\cmsRight) event samples,
in the mentioned \pt bins.
The total fit function (blue), 
the sum of the two CB functions and (only in the \jpsi case) the Gaussian function (red),
and the background continuum (black) are also shown.}
\label{fig:PR-mass-fits}
\end{figure}

\begin{figure}[ht!]
\centering
\includegraphics[width=0.95\cmsFigWidth]{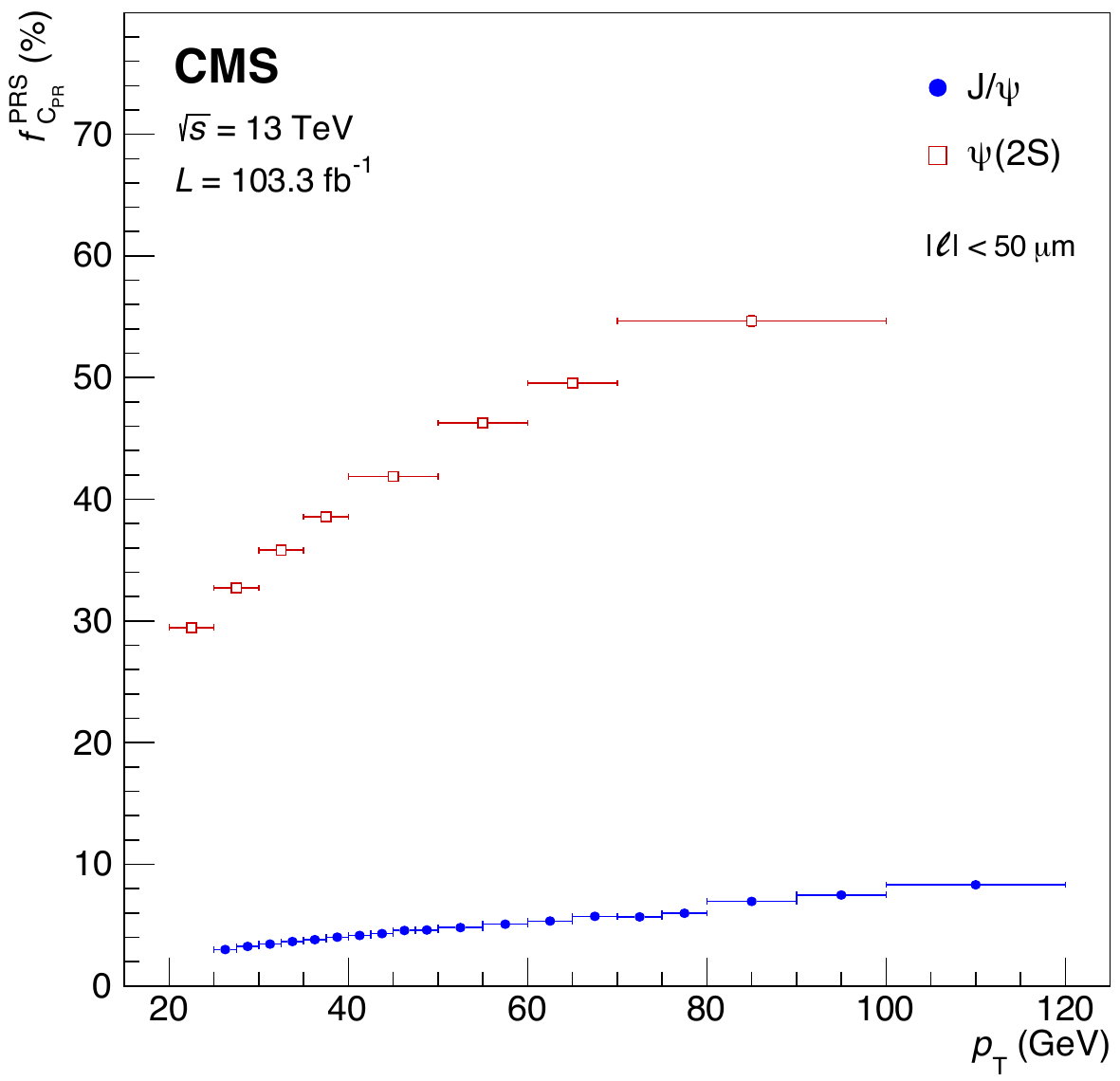}
\caption{Fraction of events in the PRS region due to continuum muon pairs,
versus \pt, for the prompt \jpsi and \psip events.}
\label{fig:PR_psi_Bg-fraction}
\end{figure}

The fractions of continuum muon pairs in the prompt signal windows,
\fPRSCPR, for both the \jpsi and \psip cases,
are determined by fitting the dimuon mass distributions in the PR region 
($\abs{\ell} < 50\,\mu$m)
using the same fit procedure and fit models as for the non-prompt cases;
the only exception is that, now also in the \jpsi case,
the two CB functions have the same $\mu$ parameter.
Figure~\ref{fig:PR-mass-fits} shows
the dimuon mass distributions measured for the PR \jpsi and \psip events,
in representative \pt bins,
while Fig.~\ref{fig:PR_psi_Bg-fraction} shows the \pt dependence
of the background fraction \fPRSCPR.

The other source of background contributing to the PRS region corresponds to 
charmonia produced in decays of short-lived \Pb hadrons.
The fractions of events in the PRS windows due to those non-prompt charmonia,
\fPRSpsiB, 
are obtained by fitting the decay length distributions of the dimuons in the 
\jpsi or \psip signal mass windows,
while taking into consideration the existence of non-prompt continuum muon pairs.
We simultaneously fit the 19 \jpsi or 8 \psip
dimuon decay length distributions (one per \pt bin),
in the range from $-50$ to $+500\mum$,
selecting the events with dimuon mass in the mass signal windows.
The fit model is composed of three contributions.
First we have the prompt term, which is represented by the decay length resolution function,
parameterized by the sum of three Gaussian functions 
with a common mean and independent widths,
defined such that $\sigma_{G3} > \sigma_{G2} > \sigma_{G1}$.
Studies of MC event samples show that the superposition of three Gaussian functions
provides a good description of the prompt decay length distributions. 
They also show that the relative contributions of the three terms, 
their mean value $\mu$, as well as the ratios 
$\sigma_{G2} / \sigma_{G1}$ and $\sigma_{G3} / \sigma_{G1}$,
are independent of \pt.
Therefore, we impose such constraints in the fit model, so that
the \pt-dependence of the decay length resolution is fully described by 
the $\sigma_{G1}$ parameter, which is seen to decrease as \pt increases.
The second term represents the contribution of $\psi$ mesons from \PB decays
and is parameterized by a decreasing exponential (for $\ell > 0$)
convolved with the resolution function.
The slope of this exponential function is left free in each \pt bin.

\begin{figure}[ht!]
\centering
\includegraphics[width=\cmsFigWidth]{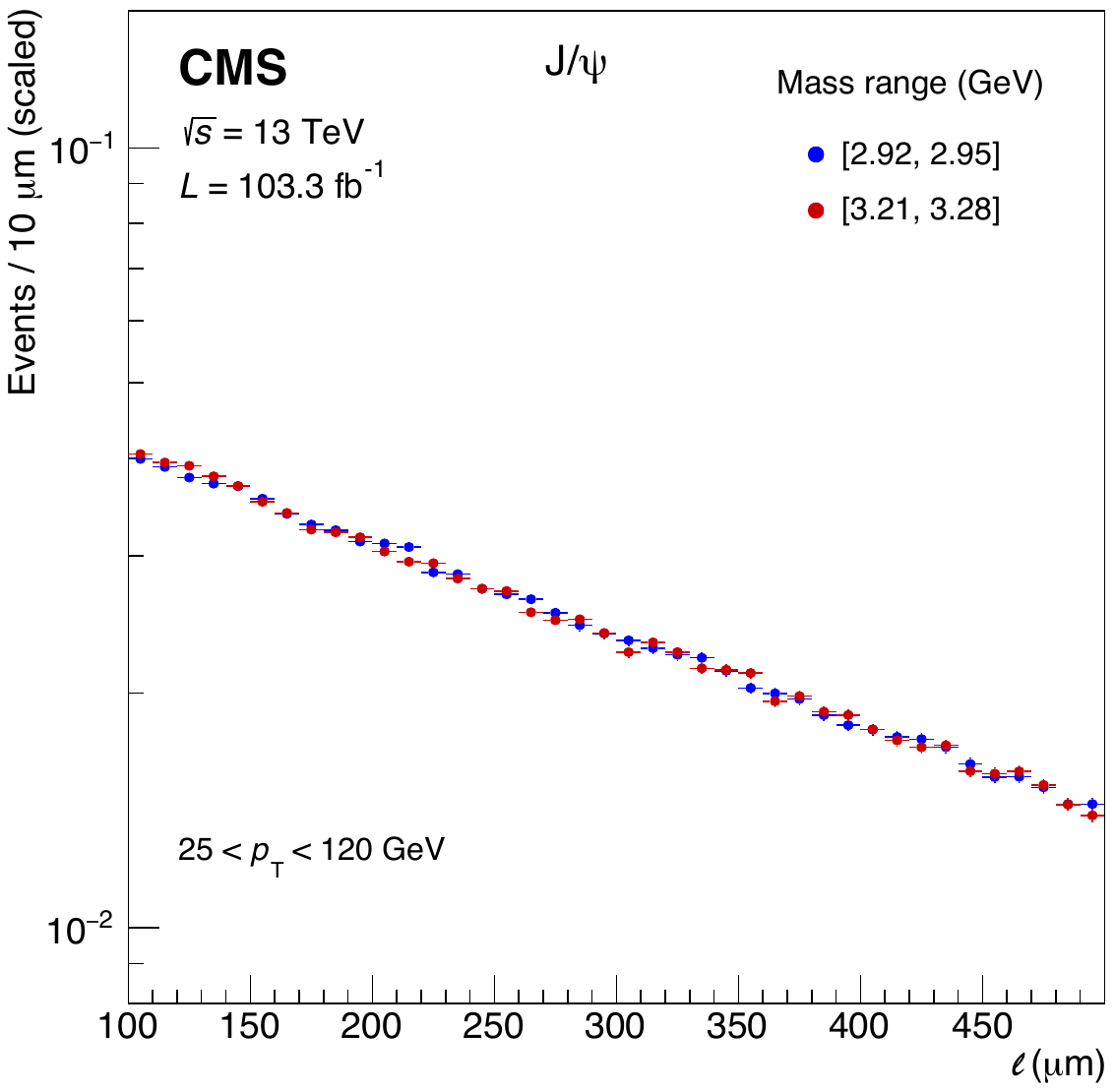}
\includegraphics[width=\cmsFigWidth]{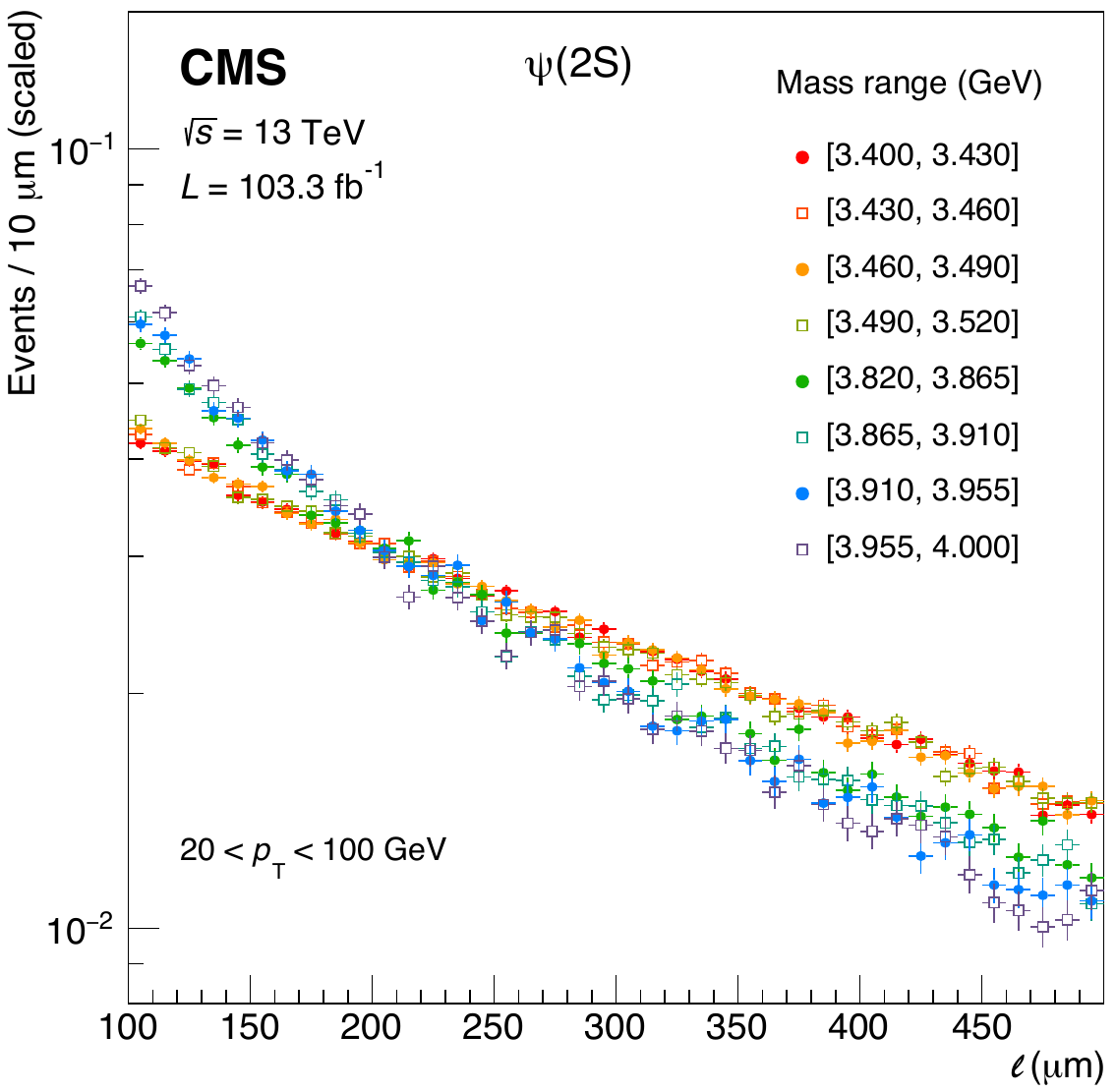}
\caption{Dimuon decay length distributions, integrated in \pt,
measured for the sideband mass ranges mentioned in the legends,
in the \jpsi (\cmsLeft) and \psip (\cmsRight) cases.}
\label{fig:lifetimes-vs-mass-sideband}
\end{figure}

The third term represents the non-prompt continuum muon pairs,
parameterized by one (for the \jpsi) or two (for the \psip) decreasing exponential functions
convolved with the resolution function.
This term is fixed, both in shape and normalization, 
by interpolating to the signal mass window the
decay length distributions (in the 100--500\mum range)
measured in the mass sidebands.
The two \jpsi mass sidebands have identical decay length distributions, 
as seen in Fig.~\ref{fig:lifetimes-vs-mass-sideband}~(\cmsLeft),
making the interpolation to the signal window a straightforward operation.
In the case of the \psip analysis,
the decay length distributions of the sideband events vary with dimuon mass.
As shown in Fig.~\ref{fig:lifetimes-vs-mass-sideband}~(\cmsRight),
the highly populated and wide mass sidebands
allow us to study the variations of the decay length distribution with mass
using four mass bins in each of the two \psip sidebands.
After finding that the shapes of the mass-sideband \psip decay length distributions
do not show any \pt dependence, 
we imposed that the functional form representing 
the decay length distribution of this background term is independent of \pt.

\begin{figure}[ht!]
\centering
\includegraphics[width=\cmsFigWidth]{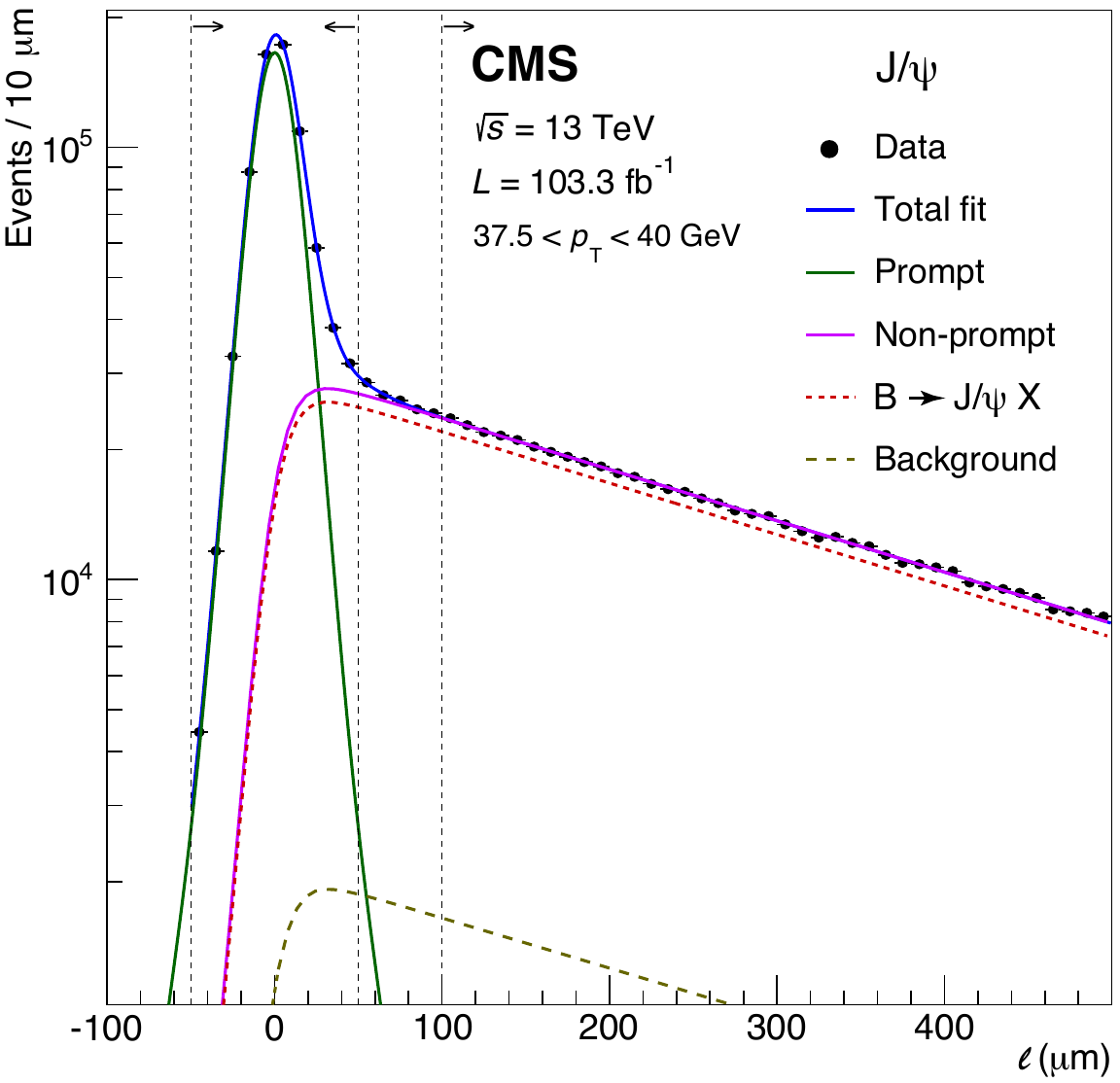}
\includegraphics[width=\cmsFigWidth]{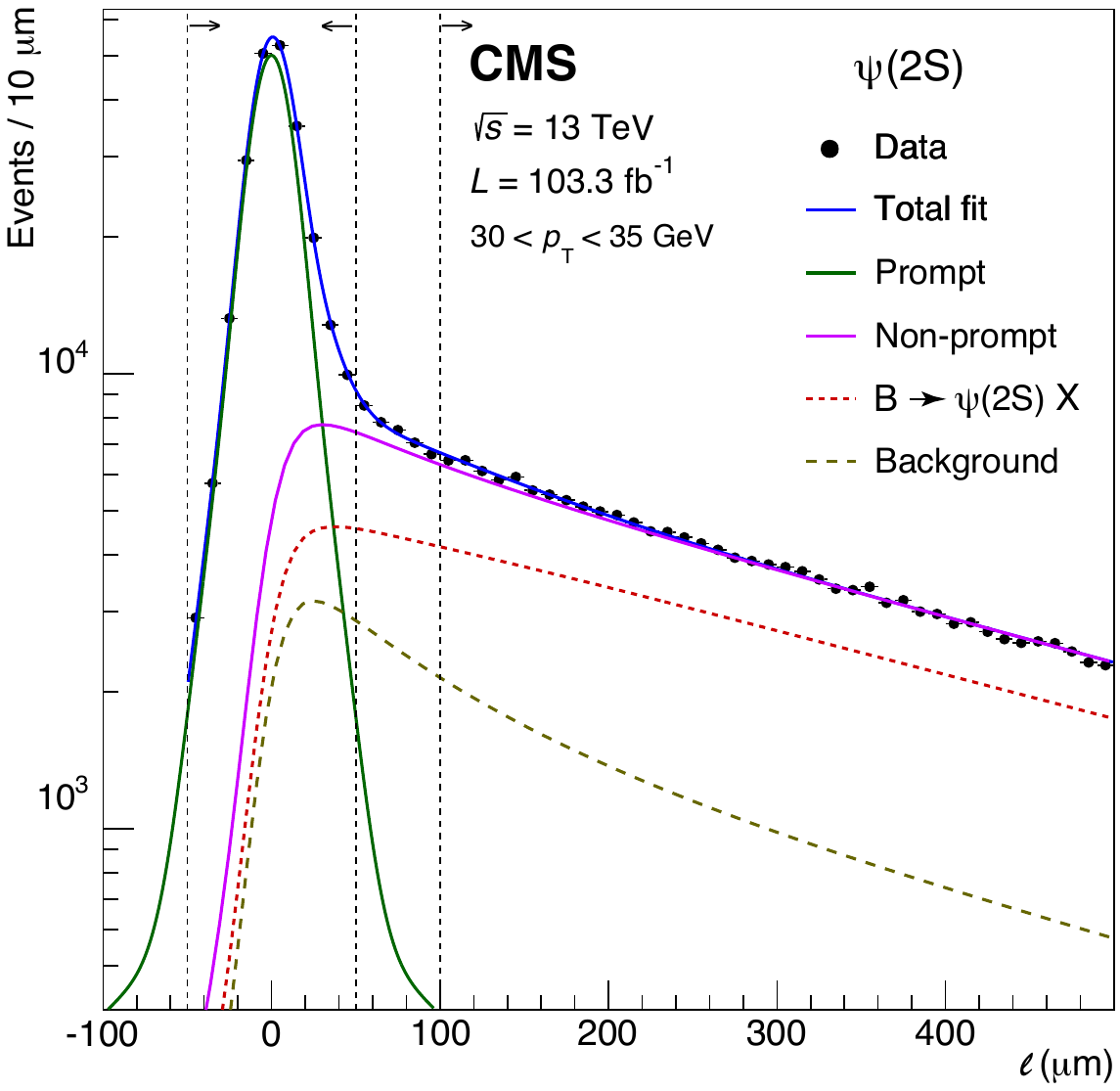}
\caption{Dimuon decay length distributions measured for the 
\jpsi (\cmsLeft) and \psip (\cmsRight) cases, 
in their mass signal windows, 
in the mentioned \pt bins.
The vertical dashed lines mark the limits of the PR and NP ranges. 
The total fit function, as well as the individual contributions, are also shown.}
\label{fig:lifetime-fits}
\end{figure}

\begin{figure}[ht!]
\centering
\includegraphics[width=0.95\cmsFigWidth]{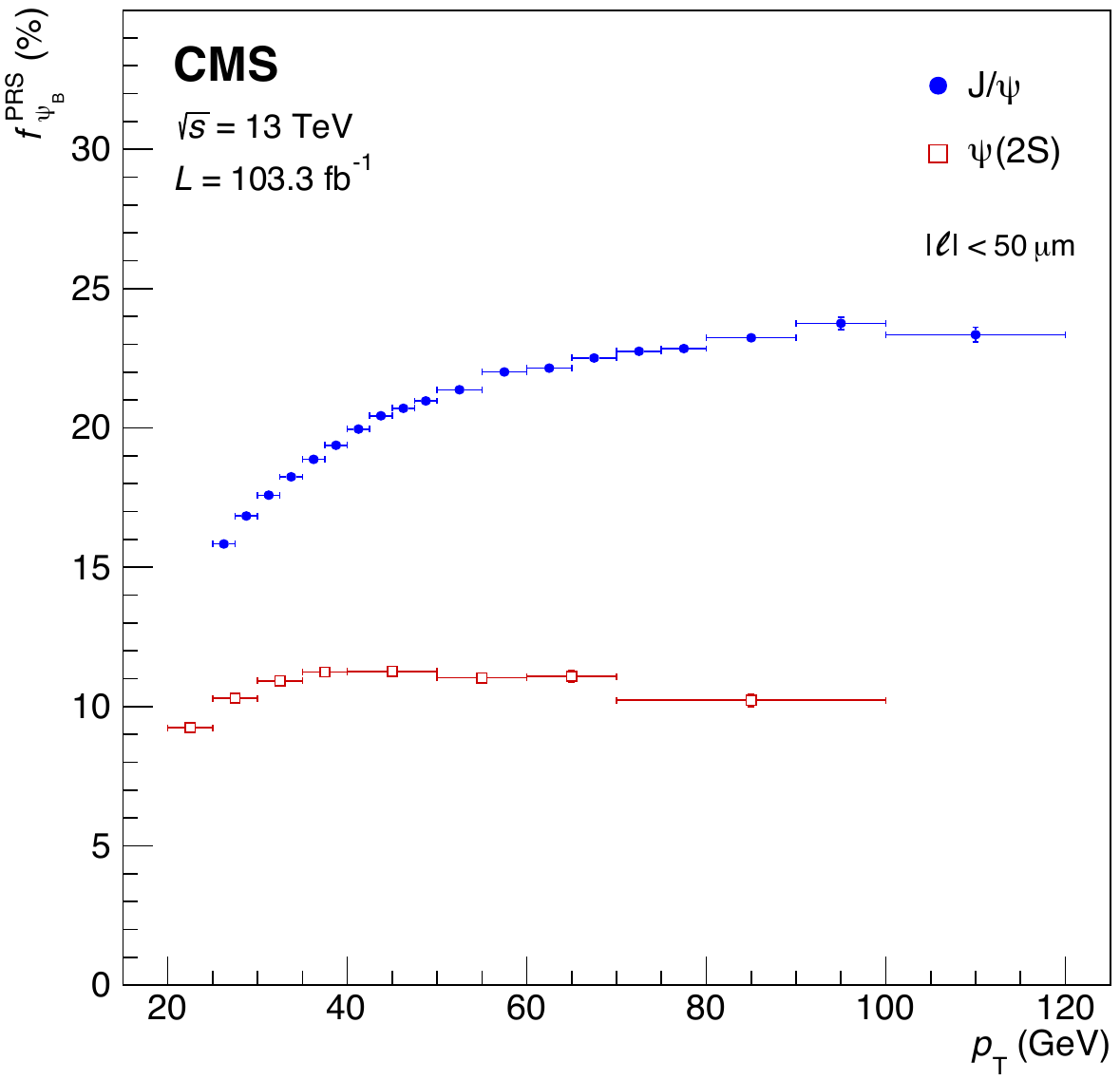}
\caption{Variation with \pt 
of the fraction of events in the PRS region from 
non-prompt \jpsi and \psip mesons.}
\label{fig:ctau-fractions}
\end{figure}

Figure~\ref{fig:lifetime-fits} shows
the dimuon decay length distributions measured in representative \pt bins for both charmonia, 
in the signal mass regions.
The \fPRSpsiB fractions are evaluated (in each \pt bin) 
by integrating the fitted \psiB term in the prompt window
and dividing the result by the total number of events counted in that region.
The obtained fractions are shown in Fig.~\ref{fig:ctau-fractions}, 
for the \jpsi and \psip cases.

\section{Polarization measurement}

\begin{figure}[ht!]
\centering
\includegraphics[width=\cmsFigWidth]{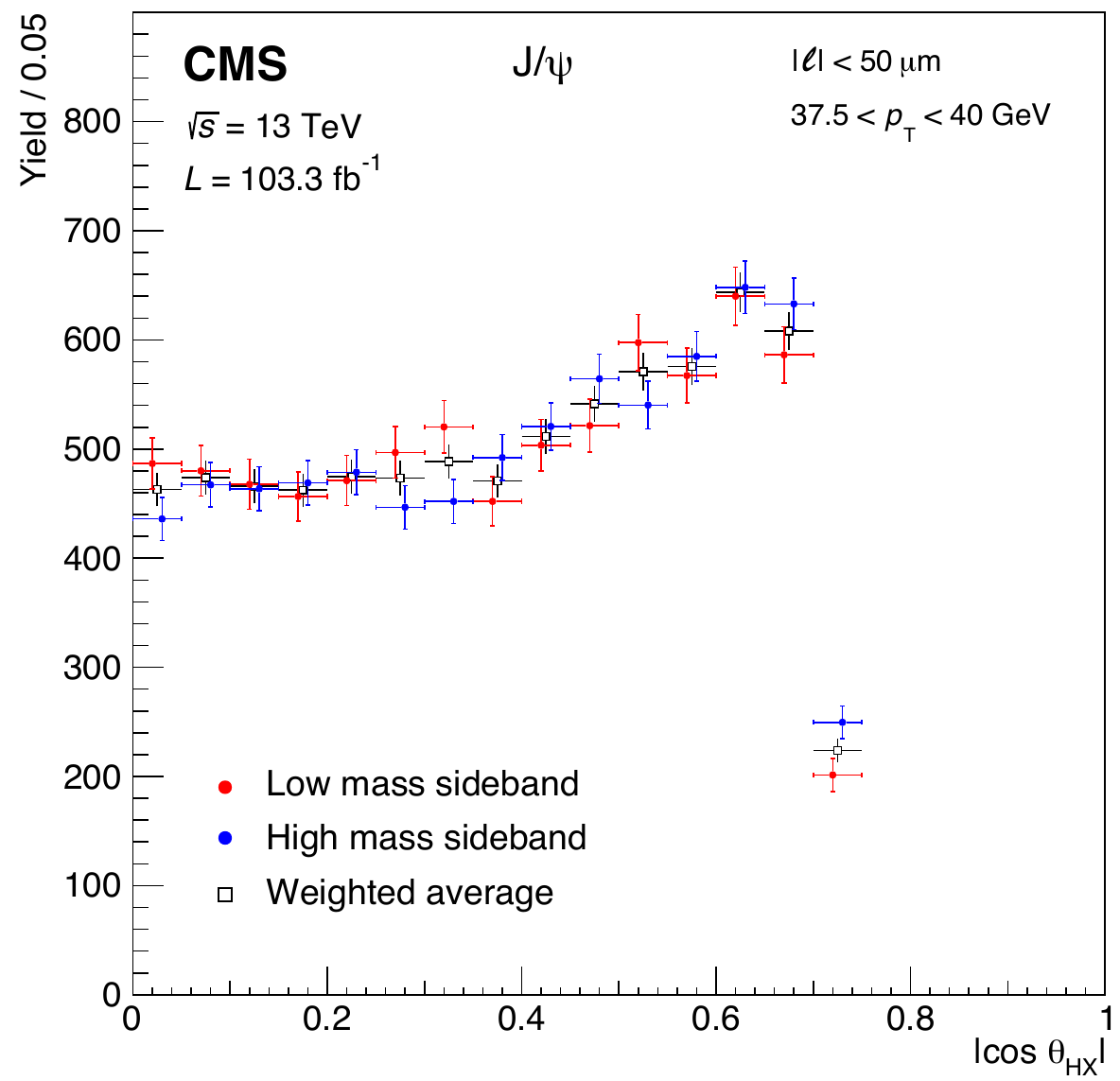}
\includegraphics[width=\cmsFigWidth]{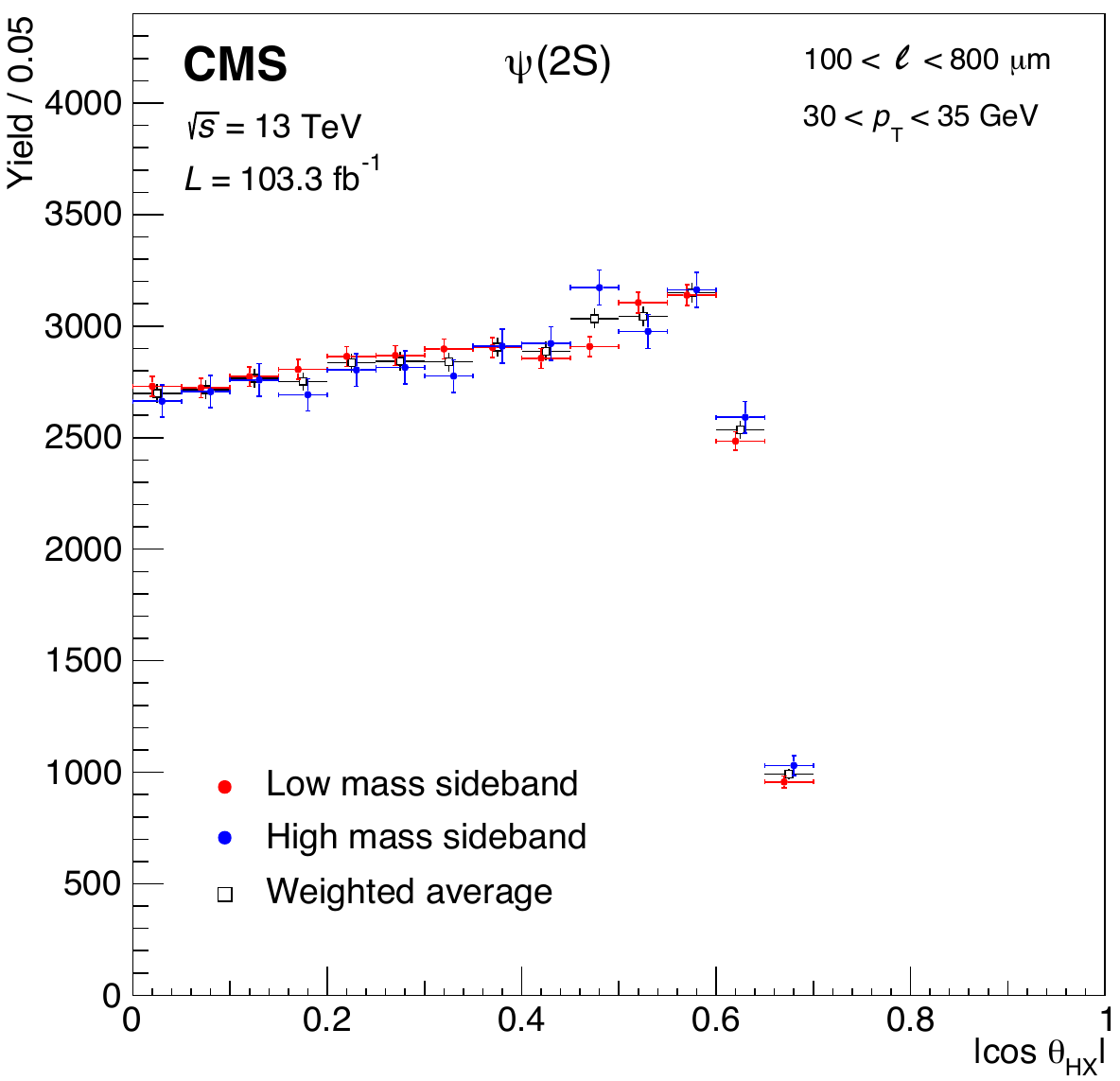}
\caption{\abscosth distributions measured in the mass sidebands
(shifted horizontally for better visibility),
and their weighted average, for the 
prompt \jpsi (\cmsLeft) and non-prompt \psip (\cmsRight) samples,
in the mentioned \pt bins.}
\label{fig:costheta-Bg}
\end{figure}

For each \pt bin,
the PRS and NPS \abscosth distributions are directly obtained from the data, 
while the \CPR and \CNP distributions are evaluated as 
weighted averages of the continuum background events falling in the low and high mass sidebands,
with weights determined by integrating the fitted 
dimuon mass background function in those sideband intervals
(seen to be essentially independent of \pt and close to 50\%, for both states).
Figure~\ref{fig:costheta-Bg} shows the \abscosth distributions 
measured in each of the two sidebands, plus their weighted average,
for two representative \pt bins of the prompt \jpsi and non-prompt \psip samples.

The non-prompt \jpsi and \psip \abscosth distributions are obtained, for each \pt bin,
by subtracting from the NPS sample the non-prompt mass continuum background,
as represented by Eq.~(\ref{eq:psi_NP}), 
using the \abscosth distributions interpolated from the mass sidebands
scaled by the previously mentioned background fractions, $f_\mathrm{\CNP}$.
An analogous procedure is followed to measure the polarizations of 
the prompt \jpsi and \psip mesons, 
the only difference being the extra subtraction of the charmonia produced in decays of \PB mesons
that contaminate the PRS window, as represented by Eq.~(\ref{eq:psi_PR}).

\begin{figure}[ht!]
\centering
\includegraphics[width=\cmsFigWidth]{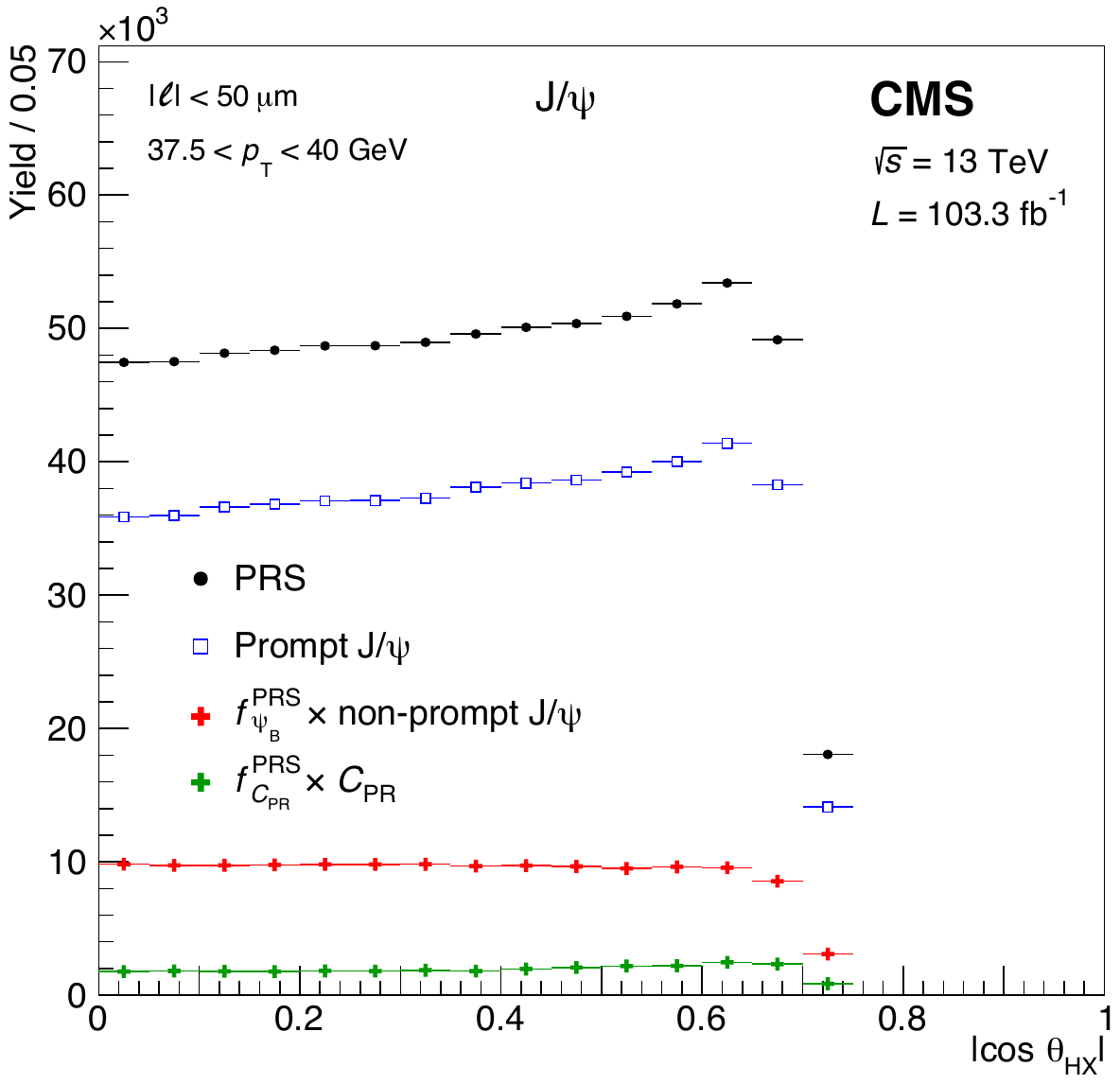}
\includegraphics[width=\cmsFigWidth]{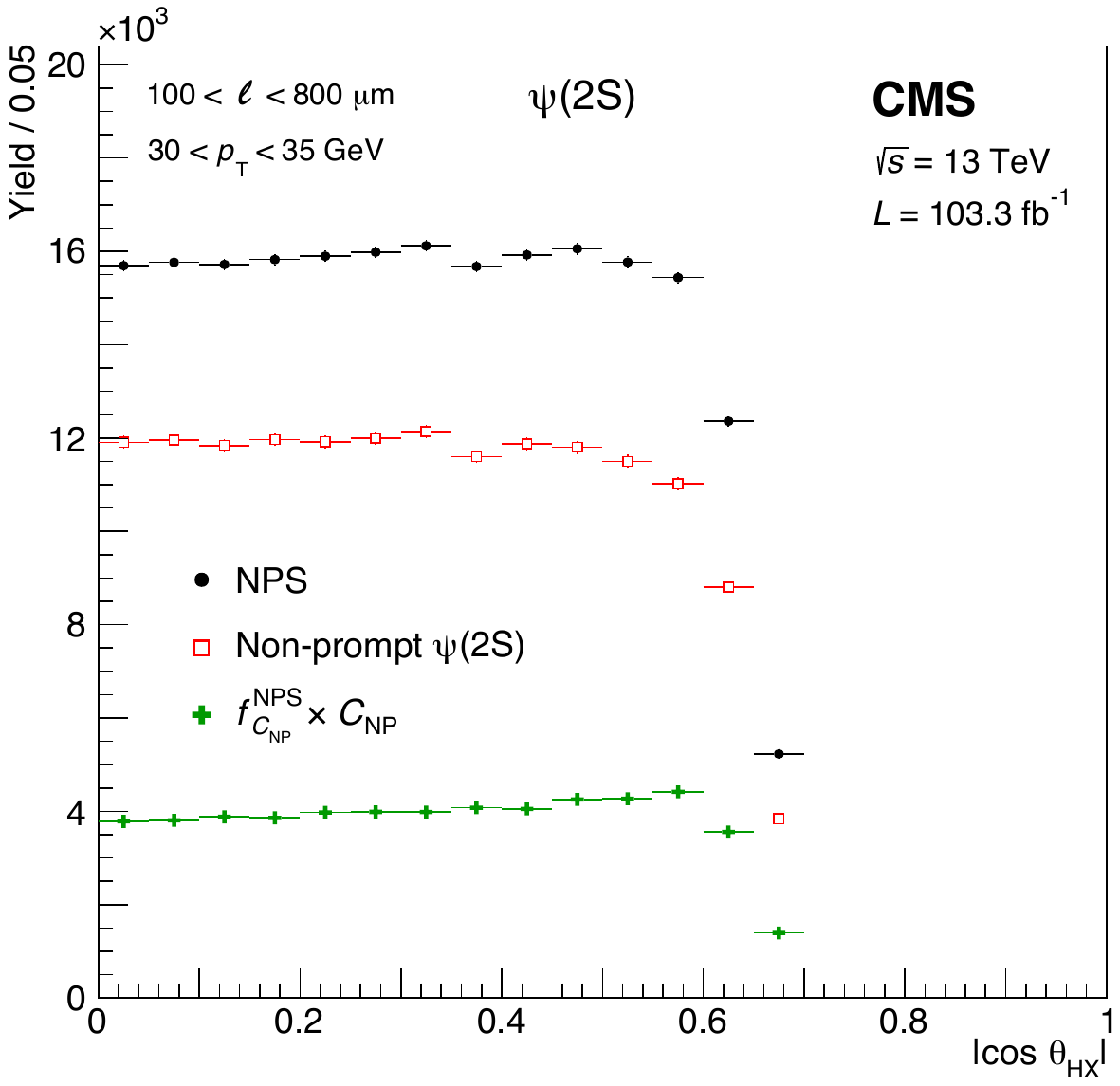}
\caption{\abscosth distributions measured in the 
PRS \jpsi (\cmsLeft) and NPS \psip (\cmsRight) samples,
of the terms of Eqs.~(\ref{eq:psi_PR}) and~(\ref{eq:psi_NP}), respectively,
in the mentioned \pt bins.}
\protect\label{fig:psi-costh}
\end{figure}

Figure~\ref{fig:psi-costh}~(\cmsLeft) shows, 
for the \jpsi events and a representative \pt bin,
the \abscosth distributions of the PRS sample (in black), 
the \psiB and \CPR contaminations
(respectively in red and green, scaled by their fractions),
and their difference, the prompt \jpsi signal (in blue).
Figure~\ref{fig:psi-costh}~(\cmsRight) shows, 
for the \psip case and an illustrative \pt bin, 
the \abscosth distributions of the NPS events (in black),
of the interpolated continuum background events (in green, scaled by its fraction),
and of their difference (in red), 
corresponding to the non-prompt \psip signal.

\begin{figure}[ht!]
\centering
\includegraphics[width=\cmsFigWidth]{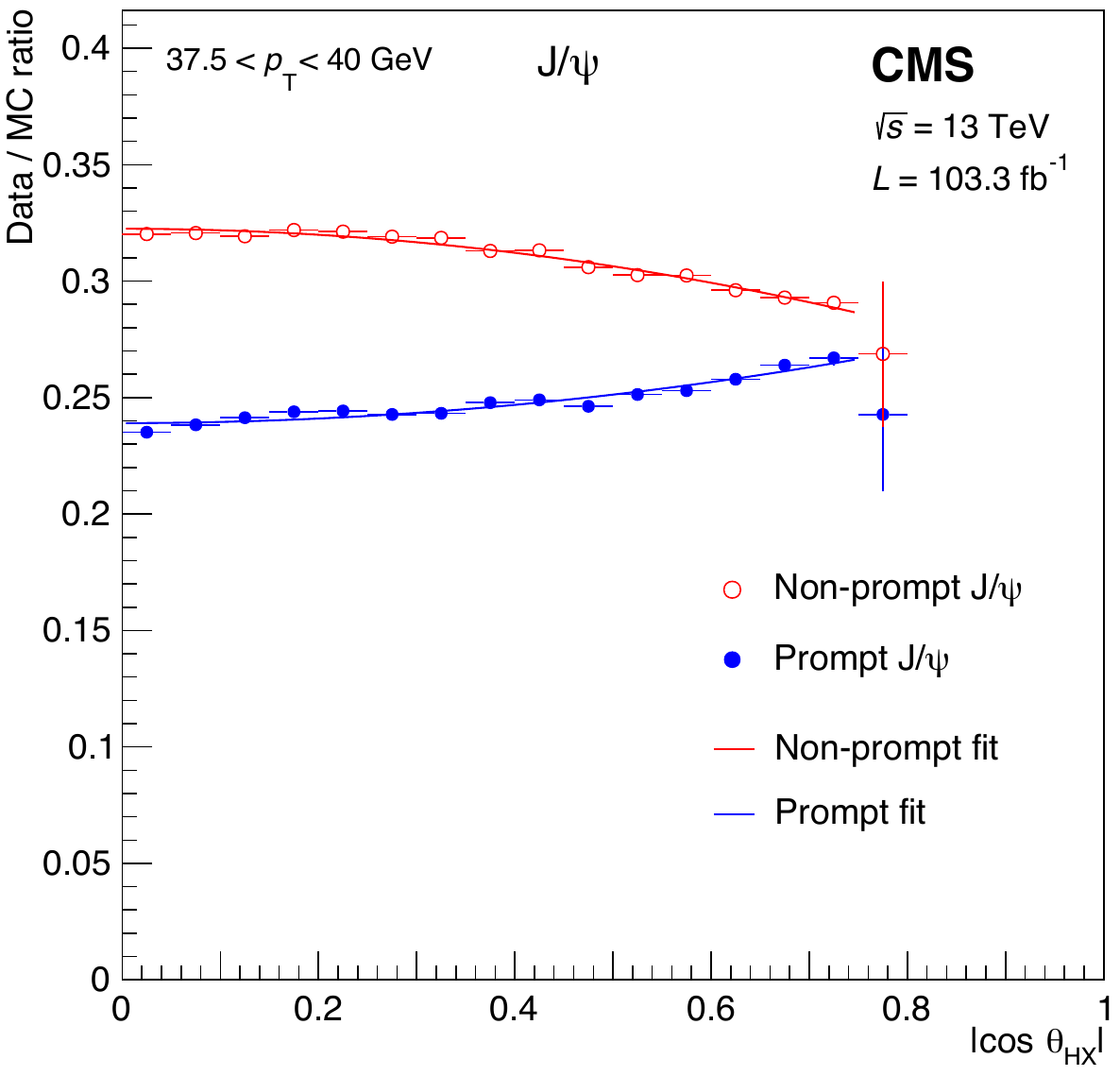}
\includegraphics[width=\cmsFigWidth]{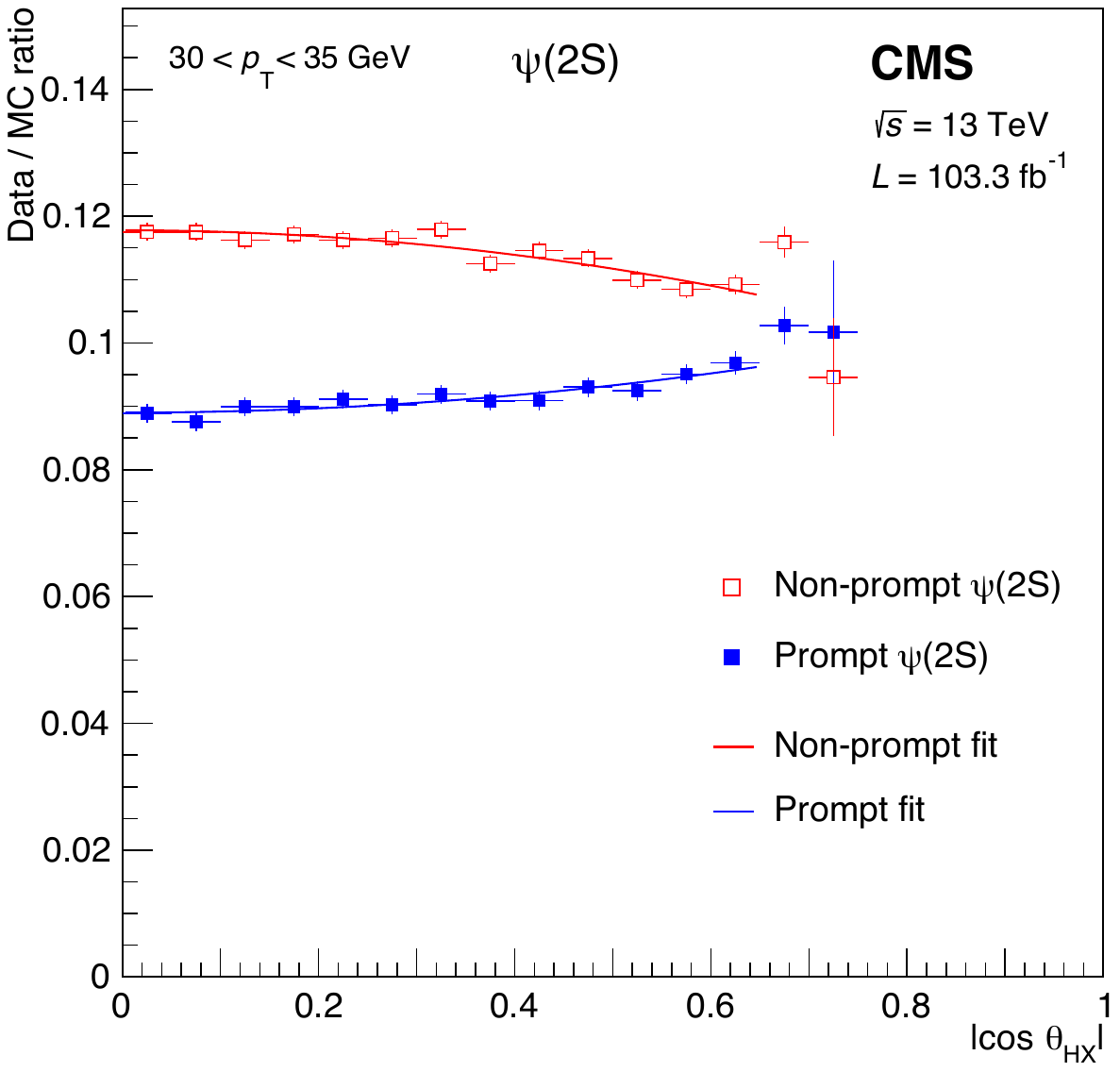}
\caption{Ratios between the measured and simulated \abscosth distributions
for the prompt and non-prompt 
\jpsi (\cmsLeft) and \psip (\cmsRight) events, 
in the \pt bins mentioned in the legends.
The curves represent fits using Eq.~(\ref{eq:W}) and 
excluding the largest \abscosth bins.}
\label{fig:psi-costh-ratio}
\end{figure}

Figure~\ref{fig:psi-costh-ratio} shows the ratios between
the measured and the simulated \abscosth distributions, 
for the prompt and non-prompt \jpsi (\cmsLeft) and \psip (\cmsRight) events,
in the same \pt bins as used in the previous figures.
Fitting these spectra with Eq.~(\ref{eq:W}) 
gives the corresponding \lth values, for this specific \pt bin.
As indicated by the curves, 
the fits do not include the \abscosth bins closest to the edge of the covered range,
corresponding to the ratio of two steeply falling distributions and 
where the uncertainties are more than twice those of the fitted \abscosth bins.
Repeating the same procedure for all \pt bins
provides the \pt-dependence of the \lth parameters, 
for the prompt and non-prompt \jpsi and \psip states.

\section{Systematic uncertainties}

The analysis has been repeated with several variations in the procedure and input parameters, 
in order to evaluate the systematic uncertainties reflecting several potential effects.

The impact of possible differences between the 2017 and 2018 event samples
has been evaluated by independently measuring the polarizations in each of the two samples.
Since the results are compatible with each other, within their (independent) statistical uncertainties,
no systematic uncertainty has been assigned.

The 5.6\GeV muon \pt threshold has been chosen to ensure that 
the selected muons have detection efficiencies
in the plateau rather than in the low-\pt ``turn-on region".
In this way the analysis is robust with respect to small differences
between the muon efficiencies in the measured and simulated event samples.
Nevertheless, some residual effects could affect the muons in the $\abs{\eta} < 0.3$ region.
Such effects have been evaluated by redoing the analysis with several independent variations:
rejecting events with at least one muon in the $0.2 < \abs{\eta} < 0.3$ region, 
where the detection efficiency is lower
because of the gap between the central muon detector wheel and its neighbours;
increasing the \pt threshold value to 6.7\GeV for muons with $\abs{\eta} < 0.2$, 
thereby completely avoiding the ``turn-on region";
and applying very conservative variations to the \pt dependence of the 
simulated muon efficiency in the $\abs{\eta} < 0.2$ region.
Only the latter variation leads to a non-negligible variation 
(restricted to the lowest \pt region) with respect to the baseline analysis.
We assign a (conservative) systematic uncertainty from this source, 
computed as the average of the absolute differences between the varied and baseline values,
decreasing from $\pm 0.011$ to zero as \pt increases from 25 to 50\GeV for the \jpsi analysis and
from $\pm 0.014$ to zero in the 20--40\GeV \pt range for the \psip case.

For $\pt \gtrsim 50$\GeV,
the two daughter muons might be emitted with 
almost parallel trajectories
and it can happen that, at the trigger level, they are detected as a single muon, 
in which case the event may not be selected~\cite{BPH-10-004,BPH-14-001}.
While this effect is expected to be reproduced by the detailed trigger emulation
included in the MC simulation, 
it is important to see if our results are sensitive to potential residual differences.
By comparing the MC event distributions before and after applying the dimuon trigger,
we can evaluate the decrease in the dimuon trigger efficiency as the two muon trajectories
approach each other in angular space (especially when they also have similar \pt values).
Such MC studies allow us to define 
criteria that select events with low dimuon trigger efficiency~\cite{Knunz:PhD},
which we have rejected from the data analysis to evaluate the sensitivity of the measurement
to a potentially inaccurate correction of this effect.
We see no differences 
(beyond fluctuations caused by the reduction in the size of the event sample)
between the baseline \lth values and those measured with the event sample 
obtained with the extra selection that rejects events with low dimuon detection efficiency.
Therefore, we assign no systematic uncertainty to cover potential inaccuracies 
in the MC dimuon trigger emulation.

The \jpsi and \psip line shapes do not enter directly 
in the determination of the fractions of continuum muon pairs 
in the signal mass windows because we use, as denominator, 
the number of counted events.
Furthermore, the signal fit model function is empirically chosen such that it
describes well the simulated and measured peak shapes.
Therefore, the dimuon mass signal shape should not have a significant impact 
on the evaluation of the continuum background fractions.
This expectation has been confirmed through explicit tests, 
as those described in the following.
In the case of the \psip analysis, 
the continuum dimuon background has a relatively high contribution,
with respect to the \psip peak, 
so that it is important, for the polarization measurement, 
to precisely evaluate the fraction of events
in the signal mass window coming from that background.
The rather broad mass sideband ranges, 
both on the left and right sides of the \psip peak,
allow us to evaluate the background fraction using an alternative procedure 
that completely avoids the necessity to describe the \psip peak:
we simply count the number of events in each of the two signal-free sideband windows 
and then compute the interpolated event yield in the signal mass window,
assuming that the mass distribution follows a decreasing exponential function.
In other words, we only consider the mass ranges where the signal peak 
has a negligible contribution, if any,
so that the background level can be computed
without the need to describe the peak line shape.
The counting method and the baseline fit procedure lead to indistinguishable results, 
confirming that the functional form used to describe the peak line shape
does not bias the extraction of the background levels,
so that we do not assign a systematic uncertainty to the \psip dimuon mass fit model.

In comparison with the \psip case, 
the low-mass sideband in the \jpsi analysis is relatively narrow 
and rather close to the left tail of the \jpsi peak,
so that a small level of signal events could contaminate the sideband window,
preventing the applicability of the counting method used in the \psip case.
Instead, we have redone the fits of the dimuon mass distributions 
with variations of the \jpsi line shape tail,
to evaluate if it could significantly contribute to the low-mass sideband window,
thereby biasing the evaluation of the level of the continuum background.
More specifically, 
we have redone the \lth measurement fixing the CB tail $\alpha$ parameter~\cite{CB}
to values equal to the baseline value plus or minus its fit uncertainty.
The varied results are indistinguishable from the baseline measurement,
so that we do not assign a systematic uncertainty reflecting the \jpsi dimuon mass fit model.

The fits of the decay length distributions include a term, 
corresponding to the non-prompt continuum dimuon background, 
that is fixed, in shape and normalization.
While the shape is well parametrized by the selected functional form
(a single exponential function for the \jpsi case 
and the sum of two exponential functions for the \psip case),
the normalization is obtained with a certain statistical uncertainty.
We have redone the prompt polarization measurement varying this normalization
by adding or subtracting its uncertainty.
The difference between the varied and baseline \lth values
is compatible with zero, independently of \pt, 
so that no systematic uncertainty is assigned to this source.

In principle, measurements of quarkonium polarizations should be made by
using Eq.~(\ref{eq:ang_distr_general}) to fit the measured two-dimensional distributions
of the polar and azimuthal decay angles~\cite{FaccioliEPJC2010}.
The one-dimensional analysis we have done, 
integrating over the \phiHX angle and using Eq.~(\ref{eq:W}),
assumes that there are no correlations between \costh and \phiHX 
in the two-dimensional acceptance maps.
This assumption is expected to be valid in the helicity frame, 
for measurements made at mid-rapidity and sufficiently high \pt,
as is the case of our analysis.
To evaluate the uncertainty in the \lth results resulting from potential residual correlations,
the analysis has been redone in exactly the same way but
replacing \abscosth by the \phiHX azimuthal angle
and fitting the acceptance-corrected distributions with the
function $1 + \beta \cos 2\phiHX$, where $\beta = (2 \, \lph) \, / \, (3+\lth)$.
As anticipated, the fitted values of $\beta$ are almost zero, 
oscillating within the $\abs{\beta} < 0.02$ range, 
for the prompt and non-prompt \jpsi and \psip mesons.
It should be noted that small non-zero $\beta$ values 
are not direct evidence of less-than-perfect MC simulations;
even an ideal simulation can lead to acceptance-corrected distributions 
that exhibit small azimuthal anisotropies.
The reason is that we are using the proton-proton HX frame 
and not the parton-parton HX frame, 
which would be the most suitable to measure the prompt polarizations.
Similarly, measurements of non-prompt polarizations would ideally be
made in the \PB meson rest frame, which is not known in our analysis.
Since we must report the measurements in the proton-proton HX frame,
small azimuthal anisotropies cannot be excluded,
even if they are absent in the ``natural frames".
Nevertheless, to be conservative, we have considered that 
the residual non-flatness of the \phiHX distributions is caused by a 
mismatch between the simulated events and the measured ones.
New \abscosth vs.\ \pt acceptance maps have been obtained, 
reweighting each MC event by the weight $1 + \beta \cos 2 \phiHX$,
using conservative $\beta$ ranges compatible with the observed azimuthal anisotropies.
Even though they are negligible with respect to the statistical uncertainty of the baseline results,
we use the differences between the \lth values obtained with the alternative maps 
and those of the baseline analysis to assign a systematic uncertainty from this source,
varying from $-0.004$ ($+0.005$) to zero in the \mbox{25--37.5}\GeV \pt range 
for the prompt (non-prompt) \jpsi polarizations and
from $-0.010$ ($+0.013$) to zero in the 20--35\GeV \pt range 
for the prompt (non-prompt) \psip polarizations.

The systematic uncertainties, summed in quadrature,
have a negligible contribution to the total uncertainties 
for $\pt \gtrsim 35$\GeV in the \jpsi results 
and for $\pt \gtrsim 30$\GeV in the \psip case.

\section{Polarization results and discussion}

\begin{figure}[ht!]
\centering
\includegraphics[width=0.48\textwidth]{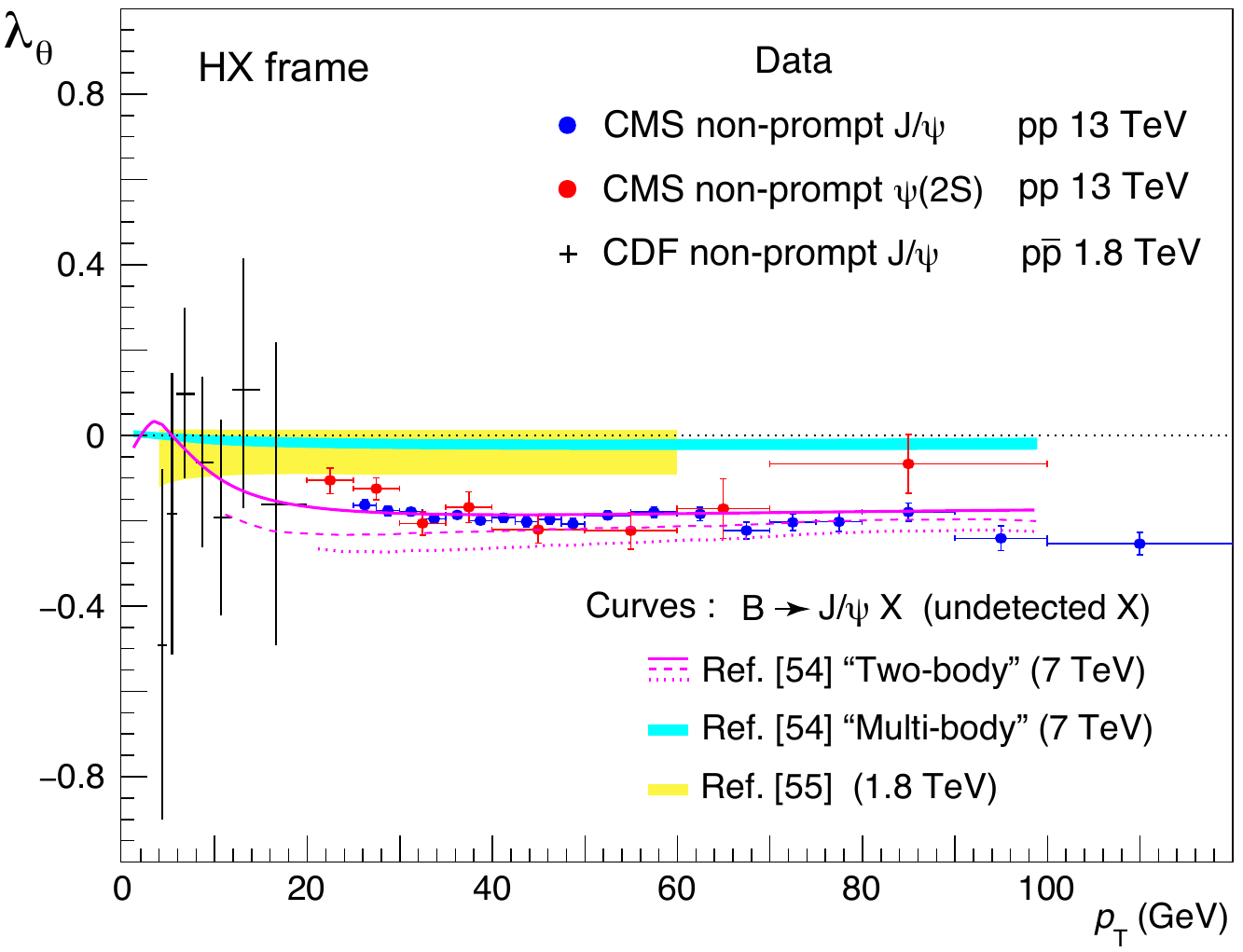}
\caption{The \lth parameter measured, as a function of \pt, 
for non-prompt \jpsi and \psip mesons.
The vertical bars represent the total uncertainties.
Predicted polarizations of \jpsi mesons produced in $\PB \to \jpsi \, X$ decays are shown 
for three calculations~\cite{NP-Pol,Krey:2002}, discussed in the text.
The low-\pt CDF measurement~\cite{CDF:nonprompt} is also shown.}
\label{fig:lthNP}
\end{figure}

Figure~\ref{fig:lthNP} shows the \lth parameter measured for the
non-prompt \jpsi and \psip mesons, as a function of \pt.
For $\pt > 30$\GeV, the measured values show a flat \pt dependence,
with \lth plateauing at around $-0.2$.
No significant differences are seen between the \jpsi and \psip trends.
The magenta and cyan bands represent the corresponding 
polarizations of \jpsi mesons produced in $\PB \to \jpsi \, X$ decays, 
computed in Ref.~\cite{NP-Pol} and briefly described in the following.
The \jpsi meson is intrinsically polarized along the direction of its emission 
in the \PB meson rest frame (the cascade helicity axis, cHX), having, \eg, 
natural polarization $\lambda_0 = -1$ when $X$ is a $J=0$ particle.
But in our ``inclusive" non-prompt measurements
we only observe the muons from the \jpsi decay, 
so that the dilepton distribution has to be referred to a direction 
(the HX axis) mostly uncorrelated from the natural cHX axis. 

Since the \jpsi is emitted isotropically in the \PB meson rest frame, 
in the \jpsi rest frame the HX and cHX directions are distributed 
in a spherically uniform way with respect to one another, 
so that one could expect
a fully smeared dilepton distribution ($\lth = 0$) in the HX frame.
However, that spherical symmetry is disrupted by the measurement process.
For example, in each \jpsi \pt bin, the $\cos \Theta$ distribution,
where $\Theta$ is the \jpsi emission angle in the \PB meson rest frame,
defined with respect to the direction of the \PB meson in the laboratory,
loses its natural uniformity and assumes a strongly modified shape, 
depending on the slope of the \pt distribution within the bin 
and on the mass of $X$~\cite{NP-Pol}. 
The predictions shown in Fig.~\ref{fig:lthNP} correspond to two hypothetical cases. 
In the one concisely denoted by ``two-body'', 
$X$ is a kaon or another $J = 0$ meson with similar or smaller mass:
$\lambda_0 = -1$ and $X$ has the kaon mass. 
The dashed and dotted curves reflect the requirement that the two \jpsi decay muons 
must have \pt values larger than 5 and 10\GeV, respectively,
while the solid curve represents the calculation without any \pt requirement.
The second, labelled as ``multi-body'', 
collectively considers all other systems of accompanying particles 
(including single particles of $J = 1$); 
the average $\lambda_0$ (of much smaller magnitude) 
and the $X$ mass distribution were derived from 
\jpsi polarizations measured in the \PB rest frame 
by CLEO~\cite{CLEO} and BaBar~\cite{BaBar}.
Comparing the data points and the computed curves indicates that
non-prompt \jpsi production (at least for $\pt > 25$\GeV) 
is dominated by the ``two-body" topology, where, interestingly, 
the \jpsi is expected to be predominantly 
produced through colour-singlet processes~\cite{Beneke:1998ks,Beneke:1999gq}. 

The non-prompt \jpsi polarization measured by CDF~\cite{CDF:nonprompt}
in $\Pp\PAp$ collisions at $\sqrt{s} = 1.8$\TeV, in the $\abs{y} < 0.6$ range,
is also included, together with a computation, shown by the yellow band, made for the CDF conditions, 
where NRQCD colour-octet contributions are found to dominate~\cite{Fleming:1996,Krey:2002}.
This prediction, leading to an almost vanishing non-prompt polarization,
is in disagreement with our measurement.

\begin{figure}[ht!]
\centering
\includegraphics[width=0.48\textwidth]{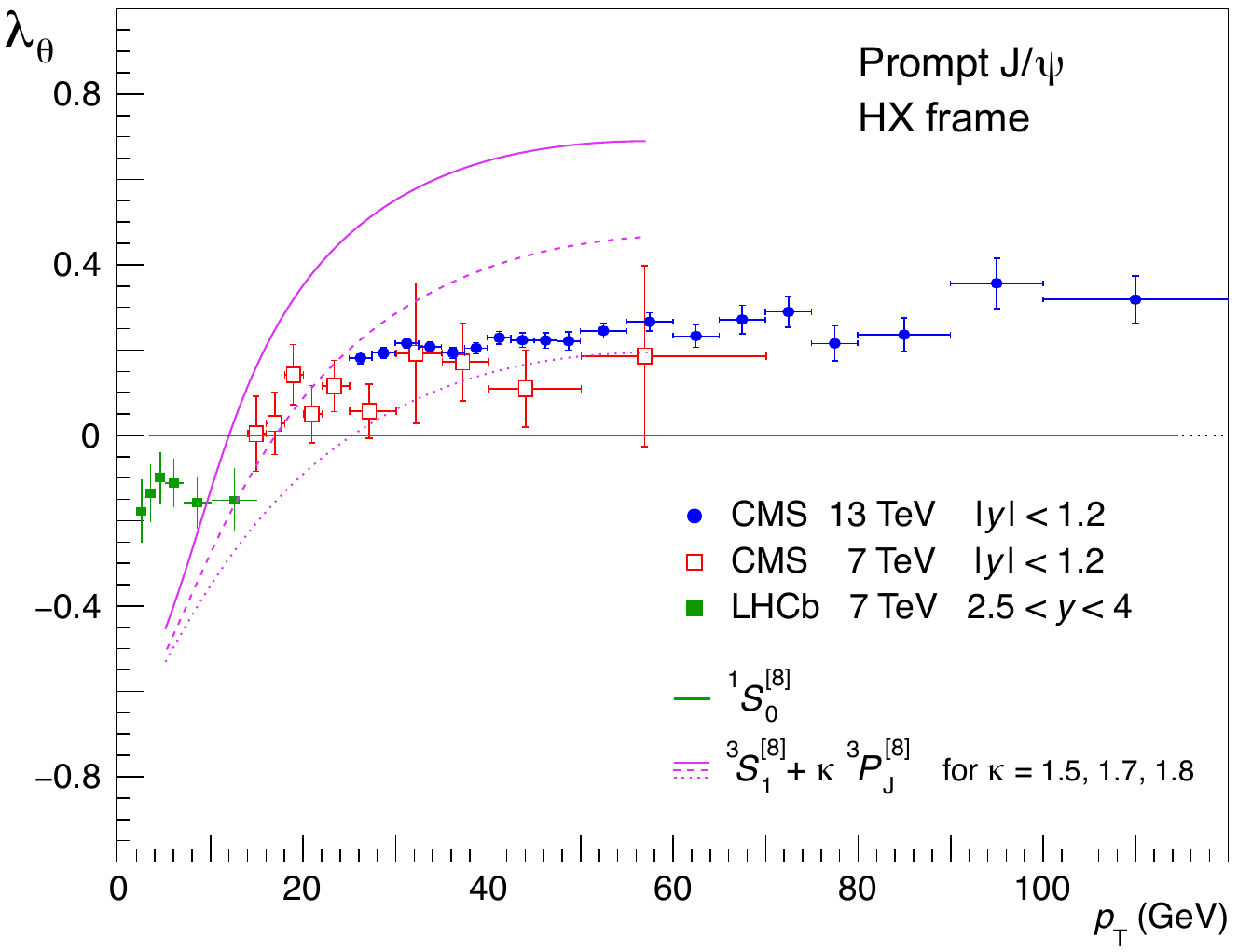}
\includegraphics[width=0.48\textwidth]{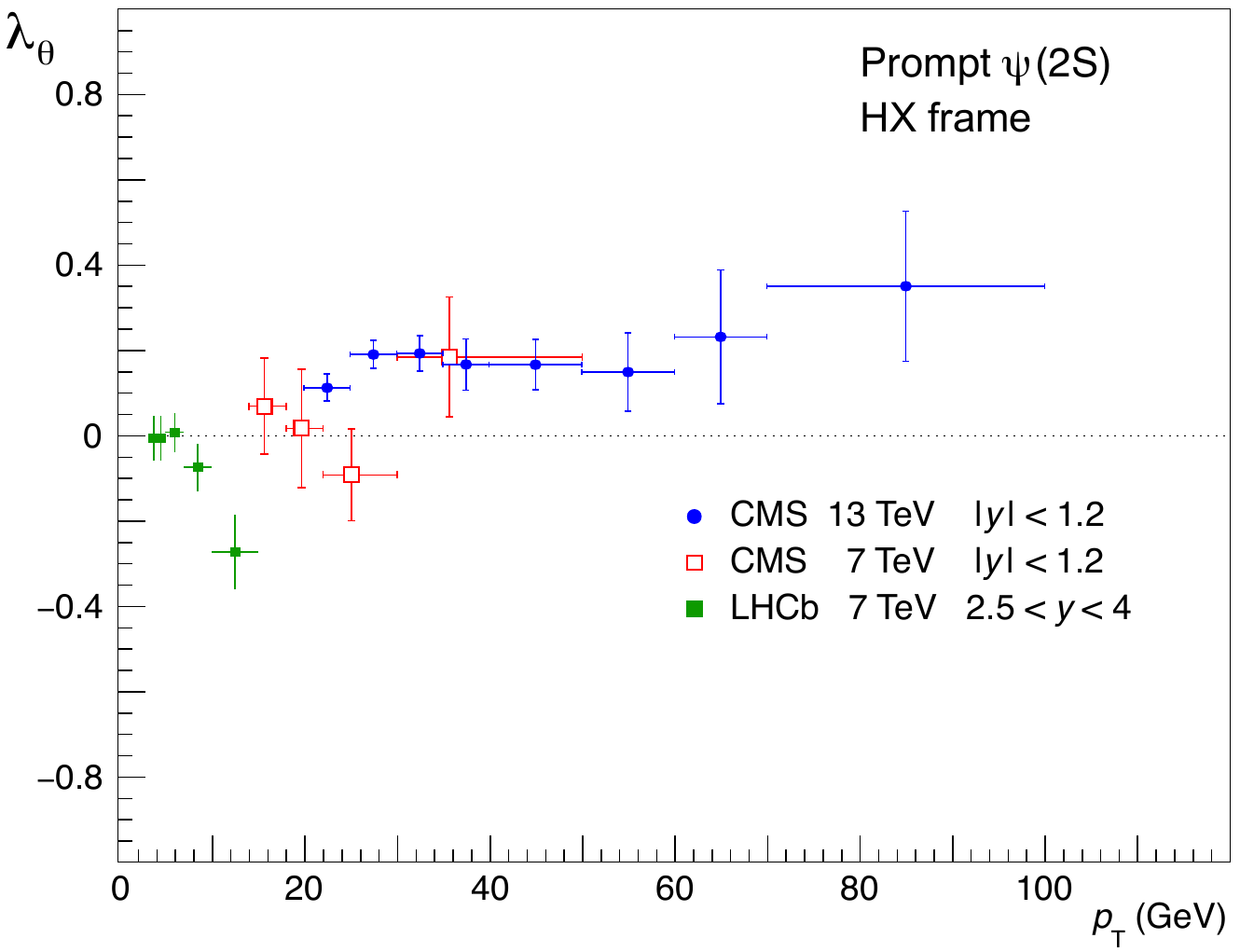}
\caption{The \lth parameter measured, as a function of \pt, 
for prompt \jpsi (\cmsLeft) and \psip (\cmsRight) mesons,
in $\Pp\Pp$ collisions at $\sqrt{s} = 13$\TeV,
compared to measurements made at 7\TeV by
CMS~\cite{BPH-13-003} and LHCb~\cite{LHCb_psiPol,LHCb_psi2SPol}.
The vertical bars represent the total uncertainties.
The curves on the \cmsLeft panel are described in the text.}
\label{fig:lthPR}
\end{figure}

Figure~\ref{fig:lthPR} shows the \lth measurements obtained in this analysis
for the prompt \jpsi (\cmsLeft) and \psip (\cmsRight) mesons.
The results previously reported by CMS and LHCb for $\Pp\Pp$ collisions at 
$\sqrt{s} = 7$\TeV~\cite{BPH-13-003,LHCb_psiPol,LHCb_psi2SPol} are also shown, 
extending the \pt coverage down to much lower \pt.
Considering all measurements together
(neglecting possible variations with collision energy and rapidity interval),
the \lth patterns show a significant trend from negative to positive, 
with the sign change occurring at $\pt \sim 15$\GeV. 
At high \pt, the polarizations tend to an asymptotic value, $\lth \approx +0.3$.
We see no evidence of the strong transverse polarizations (\lth approaching $+1$) 
expected in case the \threeSone and \threePJ octet terms would dominate.

It is interesting to compare the measured \pt dependence with the pattern obtainable 
as a superposition of the three dominant NRQCD colour octet terms, 
\oneSzero, \threeSone, and \threePJ. 
The first yields zero (\pt independent) polarization, 
while the remaining two, 
to be considered only in a mutual combination of the kind \threeSonePlusthreePJ
(the term \threePJ is unphysical when considered alone, having $\lth > +1$), 
are characterized by a \lth trend qualitatively similar to the measured one, 
going from longitudinal to transverse at a \pt transition point that depends on 
their relative proportions, symbolically represented by the $\kappa$ coefficient.
This behaviour is illustrated by the family of curves displayed in Fig.~2 of Ref.~\cite{Faccioli2019}, 
which are based on next-to-leading order calculations 
made for $\sqrt{s} =7$\TeV and mid-rapidity~\cite{Shao:2014fca,Shao:2015vga}.
Three examples of such polarized colour-octet combinations
are shown in Fig.~\ref{fig:lthPR}~(\cmsLeft), in magenta, 
together with the unpolarized term, represented by the green horizontal line at zero.
The comparison between the data points and the curves 
shows that precise polarization measurements significantly
constrain the overall NRQCD colour-octet contributions: 
the value of \pt where \lth changes sign determines $\kappa$, 
while the high-\pt asymptotic \lth value constrains the relative weight 
of the polarized (\threeSonePlusthreePJ) 
and unpolarized (\oneSzero) colour-octet terms.

The \lth values of the non-prompt and prompt \jpsi and \psip mesons 
reported in this Letter 
(corresponding to Figs.~\ref{fig:lthNP} and~\ref{fig:lthPR}) 
are provided in Tables~\ref{tab:lth-jpsi} and~\ref{tab:lth-psip},
as well as in the HEPData record for this analysis~\cite{hepdata}.

\begin{table}[ht!]
\centering
\topcaption{The \lth values measured in the HX frame, in bins of \pt, 
for the prompt and non-prompt \jpsi mesons, 
including the statistical (stat), systematic (syst), and total uncertainties.}
\label{tab:lth-jpsi}
\cmsTable{
\begin{tabular}{c|cccc|cccc}
\multirow{2}{*}{\pt (\GeVns{})} & \multicolumn{4}{c|}{Prompt \jpsi} & \multicolumn{4}{c}{Non-prompt \jpsi} \\
& \lth & stat & syst & total & \lth & stat & syst & total \\ \hline
25--27.5 & 0.182 & 0.007 & 0.012 & 0.014 & $-0.163$ & 0.006 & 0.011 & 0.013 \\
27.5--30 & 0.193 & 0.008 & 0.009 & 0.012 & $-0.176$ & 0.007 & 0.008 & 0.010 \\
30--32.5 & 0.216 & 0.009 & 0.007 & 0.012 & $-0.178$ & 0.007 & 0.007 & 0.010 \\
32.5--35 & 0.208 & 0.010 & 0.006 & 0.012 & $-0.195$ & 0.008 & 0.005 & 0.009 \\
35--37.5 & 0.193 & 0.011 & 0.005 & 0.012 & $-0.186$ & 0.009 & 0.005 & 0.010 \\
37.5--40 & 0.204 & 0.013 & 0.004 & 0.013 & $-0.200$ & 0.009 & 0.004 & 0.010 \\
40--42.5 & 0.229 & 0.014 & 0.004 & 0.015 & $-0.193$ & 0.010 & 0.003 & 0.011 \\
42.5--45 & 0.223 & 0.016 & 0.003 & 0.017 & $-0.202$ & 0.011 & 0.003 & 0.012 \\
45--47.5 & 0.223 & 0.018 & 0.002 & 0.018 & $-0.197$ & 0.012 & 0.002 & 0.012 \\
47.5--50 & 0.221 & 0.021 & 0.001 & 0.021 & $-0.207$ & 0.014 & 0.001 & 0.014 \\
50--55    & 0.245 & 0.016 & 0.000 & 0.016 & $-0.188$ & 0.010 & 0.000 & 0.010 \\
55--60    & 0.266 & 0.021 & 0.000 & 0.021 & $-0.180$ & 0.013 & 0.000 & 0.013 \\
60--65    & 0.233 & 0.026 & 0.000 & 0.026 & $-0.184$ & 0.016 & 0.000 & 0.016 \\
65--70    & 0.271 & 0.034 & 0.000 & 0.034 & $-0.223$ & 0.019 & 0.000 & 0.019 \\
70--75    & 0.289 & 0.036 & 0.000 & 0.036 & $-0.204$ & 0.019 & 0.000 & 0.019 \\
75--80    & 0.216 & 0.041 & 0.000 & 0.041 & $-0.202$ & 0.023 & 0.000 & 0.023 \\
80--90    & 0.236 & 0.039 & 0.000 & 0.039 & $-0.179$ & 0.021 & 0.000 & 0.021 \\
90--100  & 0.357 & 0.059 & 0.000 & 0.059 & $-0.241$ & 0.029 & 0.000 & 0.029 \\
100--120& 0.318 & 0.056 & 0.000 & 0.056 & $-0.254$ & 0.026 & 0.000 & 0.026 \\
\end{tabular}
}
\end{table}

\begin{table}[ht!]
\centering
\topcaption{The \lth values measured in the HX frame, as a function of \pt, 
for the prompt and non-prompt \psip mesons, 
indicating the statistical (stat), systematic (syst), and total uncertainties.}
\label{tab:lth-psip}
\cmsTable{
\begin{tabular}{c|cccc|cccc}
\multirow{2}{*}{\pt (\GeVns{})} & \multicolumn{4}{c|}{Prompt \psip} & \multicolumn{4}{c}{Non-prompt \psip} \\
& \lth & stat & syst & total & \lth & stat & syst & total \\ \hline
20--25  & 0.112 & 0.028 & 0.018 & 0.033 & $-0.105$ & 0.025 & 0.018 & 0.031 \\
25--30  & 0.191 & 0.032 & 0.010 & 0.033 & $-0.125$ & 0.024 & 0.010 & 0.026 \\
30--35  & 0.193 & 0.041 & 0.006 & 0.042 & $-0.206$ & 0.027 & 0.006 & 0.027 \\
35--40  & 0.167 & 0.060 & 0.001 & 0.060 & $-0.168$ & 0.036 & 0.001 & 0.036 \\
40--50  & 0.167 & 0.059 & 0.000 & 0.059 & $-0.221$ & 0.032 & 0.000 & 0.032 \\
50--60  & 0.149 & 0.092 & 0.000 & 0.092 & $-0.223$ & 0.043 & 0.000 & 0.043 \\
60--70  & 0.232 & 0.157 & 0.000 & 0.157 & $-0.172$ & 0.070 & 0.000 & 0.070 \\
70--100& 0.351 & 0.176 & 0.000 & 0.176 & $-0.067$ & 0.069 & 0.000 & 0.069 \\
\end{tabular}
}
\end{table}

\section{Summary}

The prompt and non-prompt \jpsi and \psip \lth polarization parameters have been measured, 
in the helicity frame and for the $\abs{y} < 1.2$ interval,
using a sample of $\Pp\Pp$ collisions at $\sqrt{s} = 13$\TeV collected in 2017 and 2018,
corresponding to an integrated luminosity of 103.3\fbinv.
The results cover \pt ranges significantly broader than previous measurements:
25--120 and 20--100\GeV, for the \jpsi and \psip, respectively.

The non-prompt \jpsi and \psip polarization measurements are compatible
with each other, regarding the \pt dependence and the overall magnitude,
plateauing at $\lth \approx -0.2$ for \mbox{$\pt > 30$\GeV}.
The measured trends agree with predictions based on the hypothesis that 
these charmonia are predominantly produced by two-body \PB meson decays,
through colour-singlet processes.

Regarding the prompt results,
we see no evidence of strong transverse polarizations (\lth approaching $+1$),
even at \pt values exceeding 30 times the \jpsi mass.
Using NRQCD concepts, there is no evidence that, at very high \pt,
the transversely polarized \threeSone and \threePJ octet terms become dominant 
with respect to the unpolarized \oneSzero octet.
Taken together with previous CMS and LHCb measurements, covering a lower \pt domain, 
we see a significant variation of the prompt polarizations with \pt, at low \pt.
These results will significantly constrain phenomenological analyses of charmonium production,
so far mostly focused on \pt-differential cross sections.

\begin{acknowledgments}
We congratulate our colleagues in the CERN accelerator departments for the excellent performance of the LHC and thank the technical and administrative staffs at CERN and at other CMS institutes for their contributions to the success of the CMS effort. In addition, we gratefully acknowledge the computing centres and personnel of the Worldwide LHC Computing Grid and other centres for delivering so effectively the computing infrastructure essential to our analyses. Finally, we acknowledge the enduring support for the construction and operation of the LHC, the CMS detector, and the supporting computing infrastructure provided by the following funding agencies: SC (Armenia), BMBWF and FWF (Austria); FNRS and FWO (Belgium); CNPq, CAPES, FAPERJ, FAPERGS, and FAPESP (Brazil); MES and BNSF (Bulgaria); CERN; CAS, MoST, and NSFC (China); MINCIENCIAS (Colombia); MSES and CSF (Croatia); RIF (Cyprus); SENESCYT (Ecuador); ERC PRG, RVTT3 and MoER TK202 (Estonia); Academy of Finland, MEC, and HIP (Finland); CEA and CNRS/IN2P3 (France); SRNSF (Georgia); BMBF, DFG, and HGF (Germany); GSRI (Greece); NKFIH (Hungary); DAE and DST (India); IPM (Iran); SFI (Ireland); INFN (Italy); MSIP and NRF (Republic of Korea); MES (Latvia); LMTLT (Lithuania); MOE and UM (Malaysia); BUAP, CINVESTAV, CONACYT, LNS, SEP, and UASLP-FAI (Mexico); MOS (Montenegro); MBIE (New Zealand); PAEC (Pakistan); MES and NSC (Poland); FCT (Portugal);  MESTD (Serbia); MCIN/AEI and PCTI (Spain); MOSTR (Sri Lanka); Swiss Funding Agencies (Switzerland); MST (Taipei); MHESI and NSTDA (Thailand); TUBITAK and TENMAK (Turkey); NASU (Ukraine); STFC (United Kingdom); DOE and NSF (USA).

\hyphenation{Rachada-pisek} Individuals have received support from the Marie-Curie programme and the European Research Council and Horizon 2020 Grant, contract Nos.\ 675440, 724704, 752730, 758316, 765710, 824093, 101115353, 101002207, and COST Action CA16108 (European Union); the Leventis Foundation; the Alfred P.\ Sloan Foundation; the Alexander von Humboldt Foundation; the Science Committee, project no. 22rl-037 (Armenia); the Belgian Federal Science Policy Office; the Fonds pour la Formation \`a la Recherche dans l'Industrie et dans l'Agriculture (FRIA-Belgium); the Agentschap voor Innovatie door Wetenschap en Technologie (IWT-Belgium); the F.R.S.-FNRS and FWO (Belgium) under the ``Excellence of Science -- EOS" -- be.h project n.\ 30820817; the Beijing Municipal Science \& Technology Commission, No. Z191100007219010 and Fundamental Research Funds for the Central Universities (China); the Ministry of Education, Youth and Sports (MEYS) of the Czech Republic; the Shota Rustaveli National Science Foundation, grant FR-22-985 (Georgia); the Deutsche Forschungsgemeinschaft (DFG), under Germany's Excellence Strategy -- EXC 2121 ``Quantum Universe" -- 390833306, and under project number 400140256 - GRK2497; the Hellenic Foundation for Research and Innovation (HFRI), Project Number 2288 (Greece); the Hungarian Academy of Sciences, the New National Excellence Program - \'UNKP, the NKFIH research grants K 131991, K 133046, K 138136, K 143460, K 143477, K 146913, K 146914, K 147048, 2020-2.2.1-ED-2021-00181, and TKP2021-NKTA-64 (Hungary); the Council of Science and Industrial Research, India; ICSC -- National Research Centre for High Performance Computing, Big Data and Quantum Computing and FAIR -- Future Artificial Intelligence Research, funded by the NextGenerationEU program (Italy); the Latvian Council of Science; the Ministry of Education and Science, project no. 2022/WK/14, and the National Science Center, contracts Opus 2021/41/B/ST2/01369 and 2021/43/B/ST2/01552 (Poland); the Funda\c{c}\~ao para a Ci\^encia e a Tecnologia, grant CEECIND/01334/2018 (Portugal); the National Priorities Research Program by Qatar National Research Fund; MCIN/AEI/10.13039/501100011033, ERDF ``a way of making Europe", and the Programa Estatal de Fomento de la Investigaci{\'o}n Cient{\'i}fica y T{\'e}cnica de Excelencia Mar\'{\i}a de Maeztu, grant MDM-2017-0765 and Programa Severo Ochoa del Principado de Asturias (Spain); the Chulalongkorn Academic into Its 2nd Century Project Advancement Project, and the National Science, Research and Innovation Fund via the Program Management Unit for Human Resources \& Institutional Development, Research and Innovation, grant B37G660013 (Thailand); the Kavli Foundation; the Nvidia Corporation; the SuperMicro Corporation; the Welch Foundation, contract C-1845; and the Weston Havens Foundation (USA).
\end{acknowledgments}

\bibliography{auto_generated}
\cleardoublepage \appendix\section{The CMS Collaboration \label{app:collab}}\begin{sloppypar}\hyphenpenalty=5000\widowpenalty=500\clubpenalty=5000
\cmsinstitute{Yerevan Physics Institute, Yerevan, Armenia}
{\tolerance=6000
A.~Hayrapetyan, A.~Tumasyan\cmsAuthorMark{1}\cmsorcid{0009-0000-0684-6742}
\par}
\cmsinstitute{Institut f\"{u}r Hochenergiephysik, Vienna, Austria}
{\tolerance=6000
W.~Adam\cmsorcid{0000-0001-9099-4341}, J.W.~Andrejkovic, T.~Bergauer\cmsorcid{0000-0002-5786-0293}, S.~Chatterjee\cmsorcid{0000-0003-2660-0349}, K.~Damanakis\cmsorcid{0000-0001-5389-2872}, M.~Dragicevic\cmsorcid{0000-0003-1967-6783}, P.S.~Hussain\cmsorcid{0000-0002-4825-5278}, M.~Jeitler\cmsAuthorMark{2}\cmsorcid{0000-0002-5141-9560}, N.~Krammer\cmsorcid{0000-0002-0548-0985}, A.~Li\cmsorcid{0000-0002-4547-116X}, D.~Liko\cmsorcid{0000-0002-3380-473X}, I.~Mikulec\cmsorcid{0000-0003-0385-2746}, J.~Schieck\cmsAuthorMark{2}\cmsorcid{0000-0002-1058-8093}, R.~Sch\"{o}fbeck\cmsorcid{0000-0002-2332-8784}, D.~Schwarz\cmsorcid{0000-0002-3821-7331}, M.~Sonawane\cmsorcid{0000-0003-0510-7010}, S.~Templ\cmsorcid{0000-0003-3137-5692}, W.~Waltenberger\cmsorcid{0000-0002-6215-7228}, C.-E.~Wulz\cmsAuthorMark{2}\cmsorcid{0000-0001-9226-5812}
% ...
\par}
\vskip\cmsinstskip
\dag:~Deceased\\
$^{1}$Also at Yerevan State University, Yerevan, Armenia\\
$^{2}$Also at TU Wien, Vienna, Austria\\
% ...
\end{sloppypar}
\end{document}